# Structure of rotator phases formed in $C_{13}$-$C_{21}$ alkanes and their mixtures: in bulk and in emulsion drops

Diana Cholakova*, Martin Pantov

Slavka Tcholakova, Nikolai Denkov

*Department of Chemical and Pharmaceutical Engineering*

*Faculty of Chemistry and Pharmacy, Sofia University,*

*1 James Bourchier Avenue, 1164 Sofia, Bulgaria*

*Corresponding authors:

Dr. Diana Cholakova

Department of Chemical and Pharmaceutical Engineering

Sofia University

1 James Bourchier Ave.,

Sofia 1164

Bulgaria

E-mail: dc@dce.uni-sofia.bg

Tel: +359 2 8161624

# ABSTRACT

Crystallization of alkane mixtures has been studied extensively for decades. However, majority of the available data consider the behaviour of alkanes with chain length of 21 C-atoms or more. Furthermore, important information about the changes of the unit cell structure with temperature is practically absent. In this work, the phase behaviour of several pure alkanes $C_nH_{2n+2}$, with $n$ ranging between 13 and 21, and their binary, ternary or multi-component equimolar mixtures are investigated by X-ray scattering techniques. Both bulk alkanes and oil-in-water emulsions of the same alkanes were studied. The obtained results show formation of mixed rotator phases for all systems with chain length difference between the neighbouring alkanes of $\Delta n \leq 3$. Partial demixing is observed when $\Delta n = 4$, yet the main fraction of the alkane molecules arranges in a mixed rotator phase in these samples. This demixing is suppressed if an alkane with intermediate chain length is added to the mixture. Interestingly, a steep temperature dependence of the interlamellar spacing in mixed rotator phases was observed upon cooling at temperatures down to 10°C below the melting temperature of the mixture. The volumetric coefficient of thermal expansion of the rotator phases of mixed alkanes ($\alpha_V \approx 2 \times 10^{-3}$ °C$^{-1}$) is around 10 times bigger compared to that of the rotator phases of pure alkanes. The experiments performed with emulsion drops containing the same alkane mixture while stabilized by different surfactants, showed that the surfactant template also affects the final lattice spacing which is observed at low temperatures. In contrast, no such dependence was observed for drops stabilized by the same surfactant while having different initial diameters – in this case only the initial temperature of the crystallization onset was affected.

## 1. Introduction

The simplest organic molecules, linear *n*-alkanes (denoted hereafter as $C_n$), are known to have complex phase behavior. When cooled from melt, they undergo one or more phase transitions to phases with an intermediate structure, called *rotator phases* (*R*), before forming the most stable crystalline phase (*C*) at low temperature [1-8]. Molecules in rotator phases form a layered structure, while the position of the molecules within the layers are not completely fixed, *i.e.* the molecules retain some mobility to rotate/oscillate around their long axis. The number and type of the rotator phases which form upon cooling/heating of bulk single-component alkane depend primarily on the alkane chain length [4-9].

The structure and properties of the rotator phases have been studied extensively in the last century after the pioneer work of Müller [1]. Initially, the rotator phases were found to form in odd-numbered alkanes with $n \geq 17$ and in even-numbered alkanes with more than 22 C-atoms [2,4,6]. These rotator phases are thermodynamically "*stable*" and form at temperature $T_{LR}$, higher than the temperature at which the liquid-to-crystal transition, $T_{LC}$, would be observed in the absence of rotator phase. Later studies of Sirota et al. [9] showed that shorter-chain even-numbered alkanes do not crystallize directly from melt, but form "*metastable*" (for $C_{18}$ and $C_{20}$) or "*transient*" ($C_{16}$) rotator phases. Both metastable and transient rotator phases have higher Gibbs energy compared to that of the crystalline phase and form at temperature $T_{LR} < T_{LC}$. Nonetheless, while the transient rotator phases are usually observed for several seconds to a few minutes only, the metastable rotator phases form in a given temperature interval upon cooling and may exist for a much longer period (up to several hours) [8,9].

The stability of the rotator phases depends not only on the length of the alkane chain but also on the length scale of the entities studied [8]. Significantly improved stability of the rotator phases is observed when the alkanes are confined in micro- or nano-environments, for example – in polymeric microcapsules [8,10-13], porous matrices [8,14-16], or emulsion drops [17-20]. In a recent study [21], employing X-ray synchrotron technique we evidenced the formation of two distinct *metastable* rotator phases within hexadecane ($C_{16}$) droplets. These droplets were stabilized by surfactants with hydrophobic tails that were similar in length to or longer than the alkane chain. In contrast, no such phases were detected for bulk $C_{16}$ in our experiments, whereas earlier studies indicated the formation of *transient* rotator phase in bulk hexadecane in approx. 10 % of the cooling experiments [9,21]. The orthorhombic rotator phase structure observed experimentally in hexadecane emulsion droplets, stabilized by appropriate long-chain surfactants, has been reproduced recently by molecular dynamics simulations [22,23].

Alkane mixing is also known to enhance the stability of rotator phases. This effect is usually explained with a reduction in the interlayer coupling interaction, combined with an increase in the lamellar surface roughness and void volume [13,24]. These combined effects result in a significantly expanded temperature range for the rotator phase existence, $\Delta T_R = T_m(\bar{n}) - T_{RC}$, where $\bar{n} = \sum_n \phi_n n$ is the molar average chain length, $\phi_n$ is the molar fraction of chain length $n$, $T_m \equiv T_{RL}$ is the melting temperature of the mixture, and $T_{RC}$ is the temperature at which the rotator-to-crystal phase transition is observed. For example, $\Delta T_R \approx 11°C$ for $C_{17}$, whereas it becomes around 25°C for a binary equimolar mixture of two consecutive chain length alkanes and increases further to 33°C for a binary mixture with $\Delta n = 2$ [24].

The mixing behavior depends primarily on the alkane chain length difference in the mixture, denoted as $\Delta n = n_2 - n_1$. Using a simple thermodynamic analysis and Bragg-Williams approximation for the solid-solution lattice [25], Matheson and Smith established that the complete mixing into a mixed solid solution occurs when the following condition is met:

$$n_{max} < 1.224\, n_{min} - 0.411, \quad (1)$$

where $n_{max}$ and $n_{min}$ represent the number of C-atoms in the longer and shorter alkanes present within the mixture, respectively [26]. This condition is not fulfilled when the Van der Waals's length of the longer alkane exceeds the length of the "lattice cell" of the smaller partner, thus making the ideal mixing highly improbable [26]. In this case, a partial or complete phase separation is expected. We note that for alkanes with chain lengths between 14 and 20 C-atoms, this rule predicts the formation of miscible phases when the chain length difference is up to 3 C-atoms, whereas mixtures with larger $\Delta n$ are predicted to phase separate. Several alternative criteria for alkane miscibility have been proposed [27-29], however, they yield similar results for alkanes with chain lengths of interest in the present study. For $n > 25$ C-atoms, minor disparities in the predicted maximum $\Delta n$ values are observed between the different models.

Alkane mixtures have been studied extensively because such mixtures are widely used in several industrial and technological applications. For example, this topic is of great importance for the petroleum industry, where most of the problems in the production, transportation and refining of crude oil arise primarily from the precipitation of paraffin waxes (mixtures of long *n*-alkanes) from liquid oil [30-32]. Alkane mixtures are also widespread in various biological waxes, such as beeswax, candelilla wax, and cocoon wax [33-35]. These mixtures are part of the thin epicuticular layer that coats the cuticle of insects and plants, regulating their wetting properties [36-38], and are present even in the cerebrum cortex of animals as constituents of the myelin and chromatin within the nuclei of cerebral cells [35,39]. The petrolatum, widely used in cosmetics as moisturizing agent, is also an example of alkane mixtures (with *n* between 25 and 70) [40].

The phase behavior of longer alkanes has been studied also, due to their wide range of applications. Numerous phase diagrams have been developed for binary mixtures of alkanes with chain lengths from 21 to 41 C-atoms, see Figure 1a,b [7,8,24,27,41-45]. Additionally, Mondieig and co-authors published several phase diagrams for binary mixtures with shorter chain lengths [45]. Using the available experimental data from the literature, Wang et al. proposed a predictive nonrandom two liquid equation able to predict the solid-rotator phase equilibrium in binary *n*-alkane mixtures [27]. While to a lesser extent, ternary [7,31,46-48] or more complex alkane mixtures involving long chain alkanes have been also investigated, often due to their relevance to diesel fuel properties [7,47-51]. Generally, it is expected that the increase of the number of components in the mixture leads to enhanced tolerance in Δ*n* after which a phase separation shall occur. In other words, a third component with an intermediate chain length acts as a "*compatibilizer*" to promote the mixture stability [24,48].

Co-crystallizing binary alkane mixtures with $n \leq 20$ usually form rotator phase $R_\mathrm{I}$, characterized by a face-centered orthorhombic lattice (space group *Fmmm*), see Ref. [47]. A bilayer stacking sequence is observed, where the long parameter of the unit cell (*c*) is equal to two molecular lengths, see Figure 1c. Using the distance between the -CH$_3$ end-groups and an estimate for the gaps between the planes of two consecutive molecular layers, the interlamellar spacing, *d*, can be estimated as [7,49]:

$$d = \frac{c}{2} \approx 1.2724(\bar{n} - 1) + 3.1476 \text{ Å} \quad (2)$$

Usually, this formula tends to underestimate the actual interlamellar spacing as determined by experimental measurements. The excess distance observed is attributed to the inherent disorder among the layers of stacked molecules within the mixture [7,49]. According to the longitudinal disorder model proposed by Dorset [51], the differences in the molecular volumes in the mixed alkane molecules are compensated by longitudinal molecular shifts within the individual lamellae. This lamellae should remain sufficiently flat to facilitate the nucleation of subsequent lamellae. It was proposed that mixtures with low variation in chain length ($\Delta n \leq 2$) arrange using interchain mixing, whereas at bigger chain length differences, the end-chains of longer molecules are twisted and/or folded [45,48].

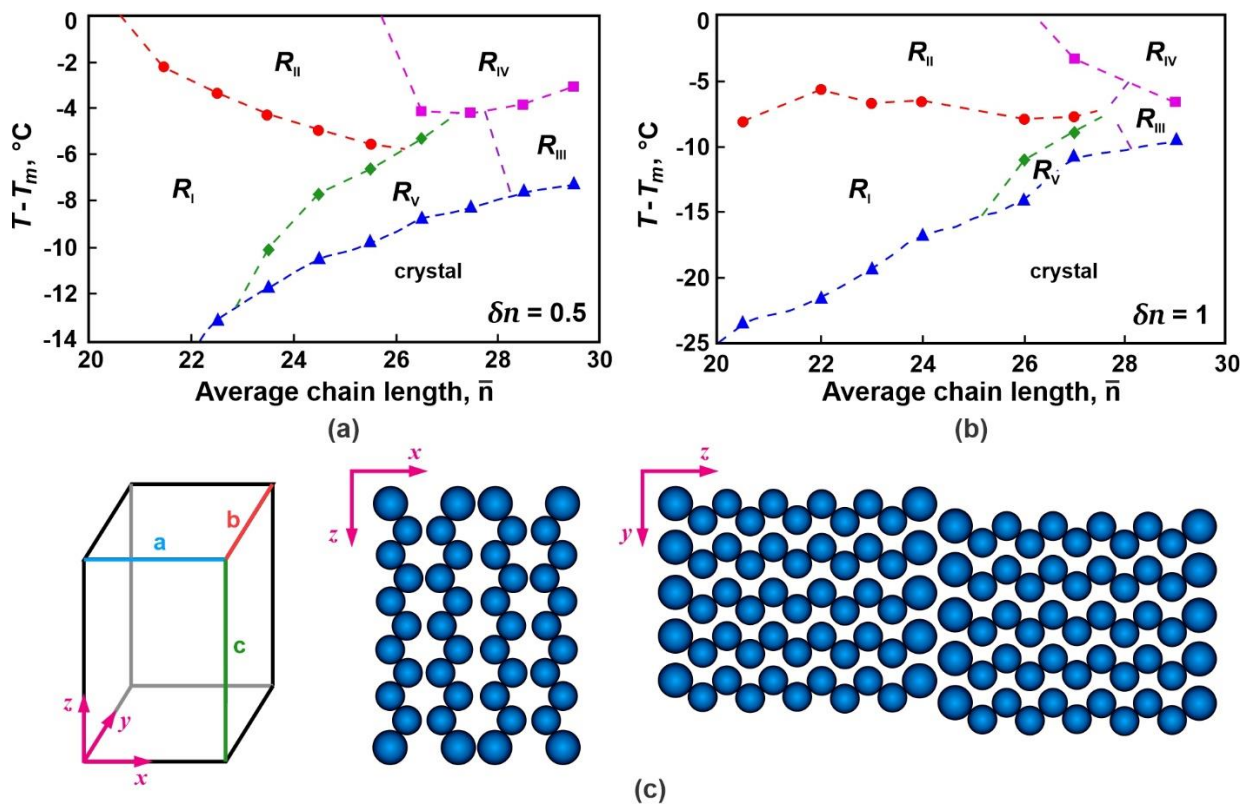

**<u>Figure 1.</u>** **(a-b)** Phase diagrams for binary alkane mixtures composed of two consecutive chain length alkanes (a) and alkanes differing by two C-atoms in chain length (b). The temperature range of existence of the rotator phase below the bulk melting temperature of the mixture is shown as a function of the average chain length. Adapted from Refs. [8,24]. **(c)** Schematic representation of the orthorhombic unit cell with its parameters *a*, *b*, *c*. The molecular arrangement with respect to the (O*xz*) and (O*zy*) planes is illustrated for odd-numbered alkanes. Note that the *c* parameter is twice the molecular length due to the *ABAB*… bilayer stacking sequence.

Despite the satisfactory level of the general understanding of the equilibrium phase behavior of alkane mixtures, the information about the structural changes occurring upon temperature variation are rather scarce. Sirota et al. [24] showed that depending on the range or rotator phase existence, $\Delta T_R$, the area per molecule jump observed upon rotator-to-crystal transition decreased from *ca.* 0.93 Å$^2$ for $\Delta T_R = 12°C$ to *ca.* 0.25 Å$^2$ when $\Delta T_R = 33°C$ (in both cases $\Delta n = 2$ and the alkane chain length was varied between 20 and 30 C-atoms). These values are significantly smaller compared to those reported for pure alkanes, where $\Delta A \approx 1.5$ Å$^2$ for $C_{17}$, $C_{19}$, or $C_{21}$ [52]. This comparison shows that the structure of the mixed rotator phases, upon temperature decrease, becomes much closer to the final crystalline structure, in contrast to the single alkane systems [24,45]. However, such data is missing for mixtures with shorter chain lengths, where the molecules are expected to be more mobile.

Furthermore, *n*-alkanes are known to exhibit the phenomenon of *surface freezing*, *i.e.* they form ordered monolayer of molecules atop molten alkanes. This monolayer can exist at temperatures up to 3°C higher than the bulk melting temperature [8,53]. Hence, it is reasonable to expect that the molecular structure may also be influenced by the surface-active components adsorbed on the emulsion drop surface. Such surfactant adsorption layers could potentially serve as templates for the crystallization of alkane molecules within the emulsion drops.

In the present study the rotator phases formed in several binary, ternary or multi-component equimolar *n*-alkane mixtures are investigated by SAXS/WAXS techniques employing both synchrotron and laboratory radiation sources. We provide also results obtained with pure alkanes for comparison. Experiments with bulk and emulsified alkane mixtures were performed to obtain both qualitative and quantitative (temperature resolved) structural information for the molecular arrangement in the studied samples. The effects of several different surfactants and of the drops size were studied for the emulsified alkanes.

## 2. Materials and methods
### 2.1 Materials

Nine *n*-alkanes and their mixtures were studied. The alkane chain lengths were varied between 13 (tridecane) and 21 (heneicosane) C-atoms. All alkanes were with purity > 99% and were purchased from TCI ($C_{13}$, $C_{15}$, $C_{19}$ and $C_{21}$) or from Sigma-Aldrich ($C_{14}$, $C_{16}$, $C_{17}$, $C_{18}$ and $C_{20}$). The alkanes were used as received.

The investigated equimolar alkane mixtures were prepared by weighting the necessary amounts of alkanes, followed by melting the mixture and its homogenization. Five binary mixtures ($C_{15} + C_{17}$, $C_{16} + C_{17}$, $C_{16} + C_{18}$, $C_{15} + C_{19}$ and $C_{17} + C_{19}$), two tertiary mixtures ($C_{15} + C_{17} + C_{19}$ and $C_{14} + C_{16} + C_{18}$) and a seven-component mixture comprising all alkanes with chain lengths between 14 and 20 C-atoms, denoted as $C_{14}$-to-$C_{20}$, were examined. The chosen mixtures had molar average chain length $\bar{n}$ between 16 and 17, except for $C_{17} + C_{19}$, where $\bar{n} = 18$. Two chain length differences were defined for mixtures having three or more mixed components: maximal chain length difference, $\Delta n_{max} = n_{max} - n_{min}$, and minimal chain length difference, $\Delta n_{min}$, defined as the smallest difference between homologous with similar chains, see Table 2 in Section 3.2.1 below.

Four different nonionic surfactants were used for stabilization of the studied alkane emulsions. Three of them were polyoxyethylene alkyl ethers, $C_nEO_m$: $C_{18}EO_{20}$ (trade name Brij S20, purchased from Sigma-Aldrich), $C_{22}EO_{30}$ (Nonidac 22-30, purchased from Sasol) and $C_{16}EO_8$ (NIKKOL BC-8SY, Nikko Chemicals). Emulsions stabilized by polyoxyethylene sorbitan monostearate $C_{16}SorbEO_{20}$ (Tween 40, Sigma-Aldrich) were also investigated. We note that Brij S20, Nonidac 22-30 and Tween 40 are commercial grade surfactants containing a broad range of ethoxy groups, with the one indicated in the chemical formula being the most abundant [54,55]. The $C_{16}EO_8$ surfactant is a specially synthesized compound, composed exclusively of this specific molecular type (purity 100% as stated by the producer). The surfactants were used as received for preparation of 1 wt. % aqueous solutions. Deionized water purified using Elix 3 module (Millipore, USA) was used, with a resistivity > 18 MΩ.cm. The selected surfactant concentration is considerably higher than the critical micellization concentration for these substances [56].

**2.2 Methods**

*Emulsions preparation and characterization*

The investigated emulsions contained monodisperse droplets, prepared by membrane emulsification technique [57]. Shirasu porous glass (SPG) membranes with pore diameters of 5, 7 and 10 µm were used in the emulsification process, yielding drops with diameters about three times bigger than the pore diameter. More detailed explanations about the emulsification process are available in our previous studies, see Refs. [21,58]. The mean volume-surface diameter of the drops, $d_{ini} = d_{32} = \sum_i d_i^3 / \sum_i d_i^2$, are used in the text. They were determined by optical microscopy observations with an AxioImager.M2m microscope (Zeiss, Germany) and Image Analysis Module of Axio Vision Software. More details about the procedure are available in Refs. [21,58].

*SAXS/WAXS (SWAXS) measurements*

The X-ray scattering measurements with *emulsified samples* were conducted in Elettra Sincotrone, Trieste, Italy, at the Austrian SAXS beamline. The experiments were performed at a fixed cooling and heating rate of 0.75°C/min. The experimental setup was similar to the one used in our previous study [21]. Briefly, 2D SAXS detector Pilatus3 1M was used simultaneously with a Pilatus 2D 100k for collection of the WAXS signal. The working energy was 8 keV ($\lambda \approx 1.54$ Å). The samples were placed in cylindrical special glass capillaries (WJM Glass, Germany) with an outer diameter of 1.5 mm, length 80 mm and wall thickness 0.01 mm. These capillaries were then placed in an aluminum thermostatic chamber connected to a thermostat. The precise temperature was measured using a calibrated thermocouple probe positioned in a neighboring orifice of the chamber. Silver behenate and p-bromobenzoic acid were used for SAXS and WAXS calibration, respectively. The correct temperature measurement was evidenced by measuring the melting points of several pure bulk alkanes, illustrative data is available in Ref. [21]. The measured melting points were always very close to those reported in the literature for the respective alkanes.

SWAXS experiments were conducted also using Xeuss 3.0 equipment (Xenocs, France) on *bulk samples* and selected emulsion samples. Capillaries with 1 mm outer diameter were used. X-rays with wavelength $\lambda \approx 1.54$ Å (Cu-K$\alpha$ radiation) were utilized, and the scattered signal was captured by an EIGER2 4M detector positioned at a sample-to-detector distance of 286 mm. Silver behenate and lanthanum hexaboride were used for detector calibration. Temperature control was achieved using the HFSX350 high-temperature stage (Linkam Scientific Instruments Ltd., UK) equipped with a T96 temperature controller and a LNP96 liquid nitrogen pump capable of achieving temperatures well below 0°C.

The spectra presented in this article have been shifted along the *y*-axis for clarity. The scattering of the sample holder has been subtracted to obtain correct intensities for all emulsion samples. The same procedure was applied to the bulk samples when necessary. However, due to the comparatively lower intensity of the X-ray source in the Xeuss apparatus compared to the synchrotron radiation source, the original curves for the bulk samples are presented in the text. The only difference between the subtracted and original curves is the presence of the Kapton peak in the latter, with a maximum at a scattering vector of $q \approx 3.26$ nm$^{-1}$. The background subtraction notably enhanced the signal-to-noise ratio in the baseline, particularly evident when the data are presented in logarithmic scale, where the secondary and ternary reflection SAXS peaks are visualized much better, without changing in any way the essential information analyzed in the paper, see Supporting Information Figure S1 for comparison. Throughout the main text, a logarithmic scale has been applied to the *y*-axis in all graphs.

To obtain the precise peak positions we analyzed the acquired signal with a Gaussian function, or performed peak deconvolution analysis involving two or more Gaussian functions in the cases of double or more complex peaks. All *d*-spacings (respectively, *c* lattice parameter values, $c = 2d$) are obtained by averaging data from the first (002), second (004), and third (006) order reflections (when available). Experimental data obtained in at least two independent experiments are used for this averaging procedure.

The *a* and *b* unit cell parameters were determined from the WAXS spectra using the relation known to be applicable for an orthorhombic unit cell [59]:

$$\frac{1}{d^2} = \frac{h^2}{a^2} + \frac{k^2}{b^2} + \frac{l^2}{c^2}, \tag{3}$$

where *h*, *k* and *l* are the Miller indices and *d* is the interplanar spacing, determined from the peak position using the relation: $d = 2\pi/q$. Note that $R_I$ phase has face-centered lattice, therefore reflections are present only when *h*, *k* and *l* have the same parity, *i.e.* they are all even or all odd numbers.

*DSC measurements*

The differential scanning calorimetry (DSC) experiments were performed on Discovery DSA 250 equipment (TA Instruments, USA). The studied sample was places in TZero aluminium hermetic pan and sealed with TZero lid using the TZero press before the measurement. The samples were cooled and heated at a fixed rate of 0.75°C/min. Before the first cooling, the samples were heated to a temperature at least 10°C higher than the melting temperature of the studied sample and maintained at this temperature for 5-10 minutes to ensure that the crystal memory of the samples had been completely erased. In preliminary experiments, we tested also longer storage periods in a molten state, but we did not observe any significant difference in the cooling/heating curves obtained after 5 min or 60 min storage in molten state. At least 3 individual cooling/heating runs were performed with each studied alkane mixture to check for the results reproducibility. Experiments were performed with some of the pure alkanes studied as well (data not shown). The observed phase transition temperatures were identical to those reported in the literature for the respective alkanes.

## 3. Experimental results and discussion

### 3.1 Phase behavior of bulk pure odd-numbered alkanes ($C_{13}$ to $C_{21}$)

This section presents the main features of the rotator and crystalline phases observed in bulk odd-numbered alkanes with chain lengths between 13 and 21 C-atoms. The even-numbered alkanes were not included in this analysis because, as explained in the introduction, the $R$ phase observed in them is very short living and difficult to observe.

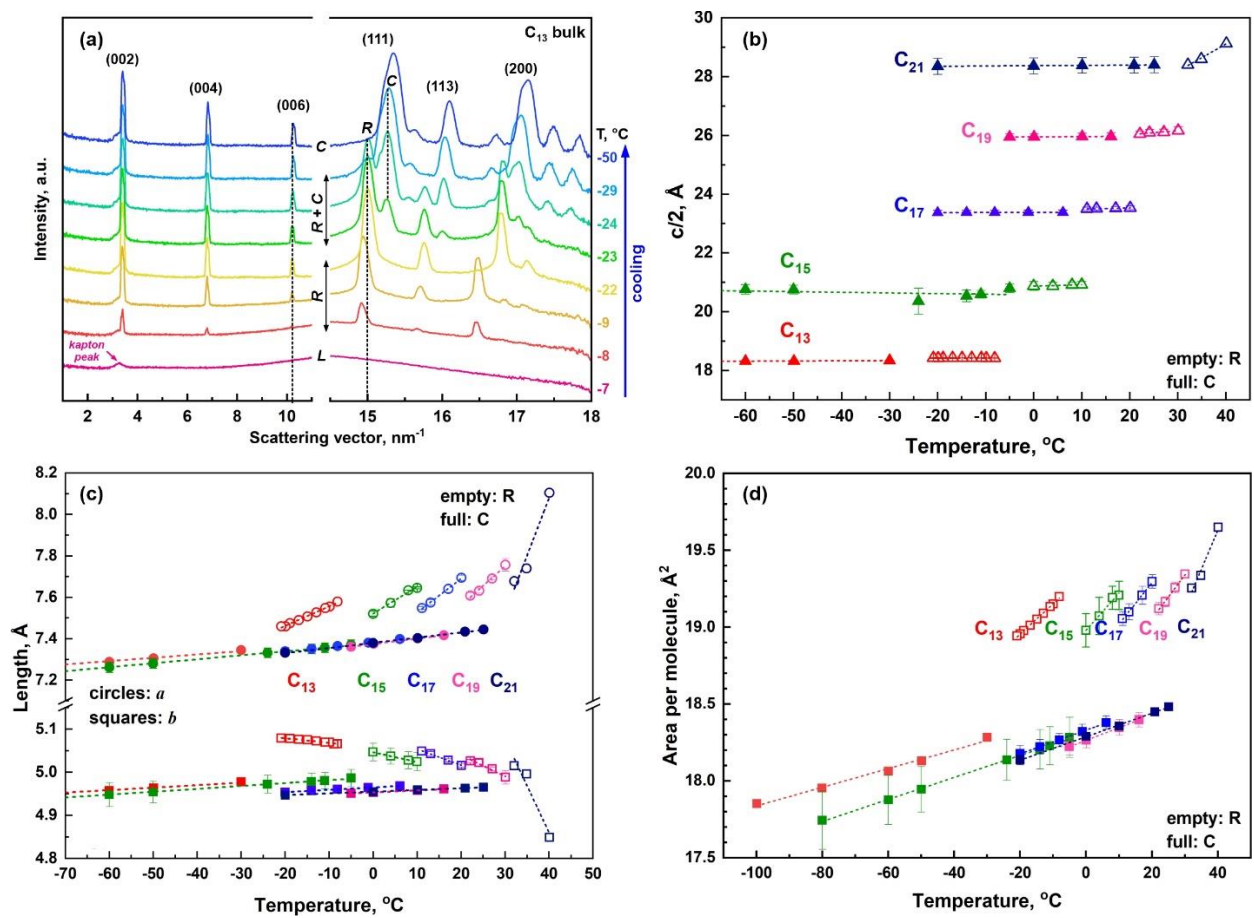

**Figure 2. Structure of bulk odd-numbered alkanes. (a)** SWAXS spectra of $C_{13}$ alkane acquired upon 0.75°C/min cooling from melt. The $L$-to-$R$ phase transition begins at $T_{LR} \approx$ -8°C. All molecules remain arranged in $R$ phase until -22°C. The beginning of the $R$-to-$C$ phase transition is detected at -23°C, see the double peaks in the WAXS spectra indicating the co-existence between the two phases. This transition ends at *ca.* -32°C. At lower temperatures the $C$ phase is observed solely. We note that the background has not been subtracted from the signal, thus the Kapton peak remains visible and slightly overlap with the primary (002) reflection. **(b-d)** Temperature dependences of: (b) interlamellar spacing $d = c/2$; (c) cell parameters $a$ (circles) and $b$ (squares); (d) area per molecule for various pure alkanes studied. The error bars in (b-d) represent the standard deviations calculated across values obtained from at least 3 independent experiments. Empty symbols show data for *R* phases and full symbols – for *C* phases. Different colors are used to distinguish different alkanes: $C_{13}$ – red; $C_{15}$ – green; $C_{17}$ – blue; $C_{19}$ – pink and $C_{21}$ – dark blue.

Figure 2a shows a typical temperature resolved SWAXS spectra for a pure odd-numbered alkane. These particular spectra were obtained in an experiment with $C_{13}$ sample. Upon cooling, the molten alkane undergoes a liquid-to-solid phase transition after which the alkane molecules arrange in *R* phase, $T_{LR} \approx -8°C$. This *R* phase is thermodynamically stable, persisting within a specific temperature range before the *R*-to-*C* phase transition begins at a lower temperature ($T_{RC} \approx -23°C$ for $C_{13}$, determined from the WAXS spectra). The first order phase transition from *R*-to-*C* phase is easily detected from the WAXS spectra due to the significant change in the area per molecule, reflected in changes of the *a* and *b* parameters in the unit cell, and associated peak shifts. Note that the position of the SAXS peaks remain relatively unchanged during this transition, as the lattice spacing remains similar. Under the applied experimental conditions, *i.e.* cooling with 0.75°C/min rate, the transition from rotator to crystalline phase is not instantaneous. Instead, the *R* and *C* phases are observed simultaneously in a certain temperature interval before the *R* phase disappears completely. In the case of $C_{13}$, the *R*-to-*C* transition ends slightly below -30°C. No further phase changes are observed when the temperature is decreased further. The observed shifts in the peak positions within the WAXS region are entirely due to the decreased thermal energy of the molecules as the temperature decreases, compare the positions of the (111) and (200) peaks at -29°C and -50°C in Figure 2a for illustration of this effect.

The acquired structural data allowed us to completely characterize the variations of the FCC lattice parameters with temperature (denoted as *a*, *b* and *c*), see red symbols in Figure 2b-d. As seen from the graphs, the lattice parameter *a* decreases as the temperature decreases for both *R* and *C* phases. Interestingly, the parameter *b* has an opposite behavior – it increases in *R* while decreasing in *C* phase upon temperature decrease. Despite the opposing behavior of *a* and *b* within the *R* phase, the overall area per molecule, denoted as *A*, diminishes while decreasing temperature, Figure 2d. This observation confirms the expected structural impact of temperature reduction. Therefore, the increase in parameter *b* as temperature decreases facilitates a faster decrease in the *a* dimension. This adjustment is necessary due to the partial rotational mobility of the alkane molecules arranged in rotator phase. Once the rotator-to-crystal phase transition occurs, both *a* and *b* dimensions begin to decrease. Furthermore, the obtained structural data shows that the slopes of the $a(T)$ and $b(T)$ linear dependences decrease after the $R$-to-$C$ phase transition. This is expected considering that the molecules in the crystalline phase exhibit substantially reduced mobility compared to those in the rotator phase.

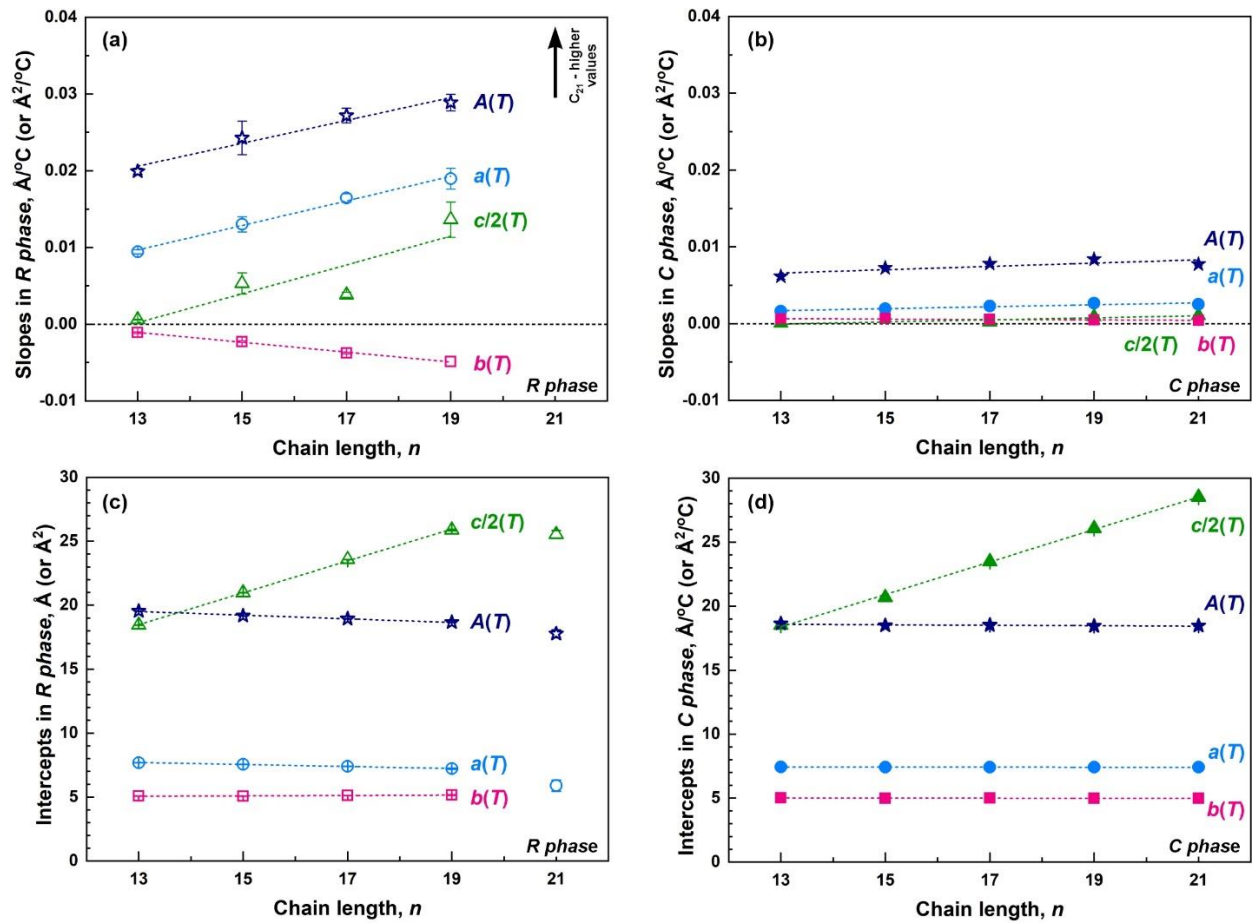

**Figure 3.** **Chain length dependences for the temperature dependent slopes (a,b) and intercepts (c,d) for functions** $f(T) = xT + y$**, where** $f \equiv a, b, c/2$ **or** $A$**.** (a,c) Rotator phases; (b,d) Crystalline phases.

Similar analyses were conducted on the other odd-numbered pure bulk alkanes with chain lengths between 15 and 21 C-atoms. The summarized results are presented in Figures 2b-d and 3, and Supporting Information Tables S1 and S2. Several important trends emerge from the obtained results: the longer hydrocarbon chain results in an increase of the temperature derivatives in $R$ phase, *viz.* the unit cell structure of longer alkanes changes faster upon unit temperature change compared to the change observed in shorter alkanes, see Figure 3a. For example, the area per molecule for $C_{13}$ changes at a rate of approximately $dA/dT \approx 0.02$ Å$^2$/°C. This value increases almost linearly reaching $\approx 0.03$ Å$^2$/°C for the $C_{19}$ alkane. The subsequent odd-numbered alkane in the homologous series, $C_{21}$, experiences a notably higher rate of change in its area per molecule with temperature, $dA/dT \approx 0.10$ Å$^2$/°C, see Supporting Information Table S1. This increase is most probably related to the increased number of structural defects within the rotator phase of $C_{21}$ alkane, *e.g. end-gauche*, *kink* or *double-gauche* conformers. These defects are known to amplify substantially with the chain length increase for *R* phases, while their number is comparable among the different chain length alkanes in the *C* phase [60,61]. Accordingly, the temperature derivatives observed in the *C* phase do not depend significantly on the alkane chain length, see Figure 3b. Therefore, the significantly higher (absolute) values of temperature derivatives for $C_{21}$ are explained with the gradual conformation changes to *all-trans* conformers as the temperature decreases.

The initial area per molecule, which can be reliably determined for *R* phase, once it is formed upon cooling from *L*, also increases when the alkane chain length increases, it is *ca.* 19.4 Å$^2$ for $C_{13}$ and 19.8 Å$^2$ for $C_{21}$, see Supporting Information Table S1. The smallest area per molecule in *R* is *ca.* 19.0 Å$^2$ for $C_{13}$ and 19.5 Å$^2$ for $C_{21}$. Similar trends are also evident for the area per molecule in crystalline phase. In this case, the absolute values for the largest possible area (as it decreases with decreasing temperature) are $\approx 18.5$ Å$^2$ for $C_{13}$ and around 18.7 Å$^2$ for $C_{21}$. All reported areas are determined with an accuracy of approximately $\pm 0.05$ Å$^2$.

The obtained results show that the *R* phase of medium chain odd-numbered alkanes can tolerate about 0.3-0.4 Å$^2$ change in the area per molecule before the *R*-to-*C* phase transition is observed. This value does not depend significantly on the alkane chain length, at least for $13 \leq n \leq 21$ odd-numbered alkanes, investigated in the current study. Furthermore, by combining the maximal $\Delta A$ found in *R* phase with the data for the temperature derivatives in different chain length alkanes, it becomes possible to explain the experimental observation that the temperature range of *R* phases existence in odd-numbered alkanes diminishes as the alkane chain length increases. For example, the temperature difference $T_m - T_{CR} = 12.6$°C for $C_{13}$ and 8.5°C for $C_{21}$ [2]. The average change in the area per molecule upon *R*-to-*C* phase transition for medium chain odd-numbered alkanes ($C_{13}$ to $C_{21}$), determined from the data obtained in the current study, is $\Delta A_{RC} \approx 0.6 \pm 0.1$ Å$^2$.

**Table 1.** Coefficients for temperature and chain length dependences of unit cell parameters and area per molecule: $f(T,n) = (gn+m)T + pn + r$. The formula is applicable to odd-numbered alkanes with chain lengths between 13 and 19 C-atoms in $R$ phase, and for $n = 13$ to 21 in the $C$ phase. This applies within the temperature range wherein a specific phase of a given alkane exists. The numbers in brackets denote the maximal standard deviations for the respective constant. The constants $g$ and $m$ are given in units of Å/°C, while $p$ and $r$ constants are expressed in Å for $a(T, n)$, $b(T, n)$ and $c/2(T, n)$ functions, whereas for $A(T, n)$ the units are Å²/°C and Å², respectively.

| $f(T,n)$ | Constants for $R$ phases | | | | Constants for $C$ phases | | | |
|---|---|---|---|---|---|---|---|---|
| | $g$ | $m$ | $p$ | $r$ | $g$ | $m$ | $p$ | $r$ |
| $a(T,n)$ | 1.6 × 10⁻² (9.1×10⁻⁵) | -1.1 × 10⁻² (1.5×10⁻³) | -7.8 × 10⁻² (4.2×10⁻³) | 8.71 (0.07) | 1.3 × 10⁻⁴ (2.8×10⁻⁵) | ≈ 0 | -1.0 × 10⁻³ (8.2×10⁻⁴) | 7.44 (0.01) |
| $b(T,n)$ | -6.4 × 10⁻⁴ (2.5×10⁻⁵) | 7.3 × 10⁻³ (4.0×10⁻⁴) | 1.4 × 10⁻² (4.9×10⁻³) | 4.88 (0.08) | -2.7 × 10⁻⁵ (9.9×10⁻⁶) | 1.0 × 10⁻³ (1.7×10⁻⁴) | -4.3 × 10⁻³ (1.5×10⁻³) | 5.06 (0.03) |
| $\frac{c}{2}(T,n)$ | 1.9 × 10⁻² (7.4×10⁻⁴) | -2.4 × 10⁻² (1.2×10⁻²) | 1.25 (0.02) | 2.29 (0.35) | 1.3 × 10⁻⁴ (3.9×10⁻⁵) | -1.8 × 10⁻³ (6.9×10⁻⁴) | 1.27 (0.03) | 1.87 (0.47) |
| $A(T,n)$ | 1.5 × 10⁻³ (2.3×10⁻⁴) | 1.5 × 10⁻³ (2.3×10⁻⁴) | -0.14 (0.01) | 21.35 (0.16) | 2.2 × 10⁻⁴ (8.9×10⁻⁵) | 3.8 × 10⁻³ (1.5×10⁻³) | -1.8 × 10⁻² (7.6×10⁻³) | 18.82 (0.13) |

Using the temperature dependent data presented in Figure 2, we derived equations that encompass both the chain length and temperature effects, *i.e.* $f(T,n)$, where $T$ is the temperature expressed in degrees Celsius, $n$ is the alkane chain length, and $f$ denotes the function of interest ($a$, $b$, $c$ or $A$). To do this, we plotted the slopes and intercepts of the linear temperature dependences determined from the experimental data as a function of the alkane chain length, see Figure 3. The resulting trends exhibited linearity for odd-numbered alkanes with chain length $13 \leq n \leq 19$ in $R$ phase and $\leq 21$ for the $C$ phase (note that alkanes of shorter or longer lengths were not studied here). Thus, by combining the derived equations with the initial linear temperature dependences, we obtained a combined expression for $f(T,n)$:

$$f(T,n) = x(n)T + y(n) = (gn+m)T + pn + r \qquad (4)$$

The constants $g$, $m$, $p$ and $r$ in equation 4 were determined for each data set, see Table 1. By using these values, one can calculate $a(T, n)$; $b(T, n)$; $c(T, n)$ and $A(T, n)$ functions for an arbitrary temperature within the temperature interval of existence of $R$ and $C$ phase for odd-numbered alkanes with chain lengths between 13 and 21 C-atoms. A comparison between the experimentally measured parameters and those calculated using the derived dependences is presented in Supporting Information Table S2. The error between the values calculated using these formulas and the experimentally obtained ones is *ca.* 1% for $a(T, n)$ and $A(T, n)$, and $\approx$ 1.5 to 2% for $b(T, n)$ and $c(T, n)$ dependences.

We note that the derived equations are linear with respect to $T$ and $n$. The linearity in the $c(n)$ dependence can be easily explained considering the fact that the long lattice parameter $c$ depends mainly on the alkane chain length, as discussed in the Introduction section, see also eq. 2 above [7,49]. The short lattice parameters (*a* and *b*) do not depend significantly on the chain length, as they show the distances between the alkane molecules in the layer, see Figure 2c. However, they depend strongly on the temperature. This dependence can be explained considering the specific thermal molecular motions. The thermal expansion is known to occur in a continuous way [62]. A theory for the thermal expansion of crystalline solids has been developed by Grüneisen [63]. According to this theory, the thermal expansion of the solids is associated with the so called 'thermal pressure' which originates from the lattice vibrations. The thermal pressure increases with the increase of temperature. In alkane systems, the anisotropic thermal pressure along the *a* and *b* axes arises from the rotational and torsional vibrations of the chains, resulting in different thermal expansions of the *a* and *b* lattice parameters. The observed thermal effects for the third lattice parameter *c*, has been explained with changes of the thickness of the plane of voids, situated between the adjacent lamellas [62]. Linear changes in the lattice parameters upon change of temperature was shown also for pentacontane sample ($C_{50}H_{102}$) via molecular dynamics simulations [64].

The newly obtained data allowed us to calculate also the thermal expansion coefficients $\alpha_f$ see Supporting Information Table S3, defined as [62]:

$$\alpha_f = \frac{1}{f_0}\left(\frac{df}{dT}\right)_P \tag{5}$$

where $f$ is either of the unit cell parameters (*a*, *b* or *c*) or the unit cell volume $V$. Here $f_0$ is the reference value of the studied parameter. The volumetric expansion coefficient for the rotator phases of pure odd-numbered alkanes with $n$ = 13-19 was $\alpha_{V,R} \approx 1.5 \times 10^{-4}$ °C$^{-1}$, whereas for the orthorhombic crystalline phases it was more than twice smaller, $\alpha_{V,C} \approx 6.3 \times 10^{-5}$ °C$^{-1}$. We note that the derived values are in a good agreement with those published in the literature, for example $\alpha_{V,C} \approx 4.2 \times 10^{-4}$ °C$^{-1}$ for the triclinic lattice of $C_{16}$ and $C_{18}$, see Ref. [65]. Similar values as the presently obtained ones are also shown for $\alpha_a$ and $\alpha_b$ in Refs. [62,66].

The results obtained for odd-numbered pure alkanes serve as a reference for discussing the results obtained with alkane mixtures, which are presented in the following section.

### 3.2 Structure of rotator phases formed in mixed alkanes

In this section, we present the experimental results obtained from equimolar binary, ternary, and more complex alkane mixtures. Experiments with both bulk and emulsified mixtures were performed, aiming to understand whether any significant difference exists between the temperature dependent rotator phases in these two types of systems. The experiments with oil-in-water emulsions allowed us to assess how the surfactant chain length and confinement size (viz. the drop size) affect the phase structures. The results are organized as follows: Section 3.2.1 shows a series of illustrative temperature resolved SWAXS spectra, elucidating the main types of phase behavior characterized in the present study; Section 3.2.2 illustrates the impact of surfactant properties and initial drop size on the rotator phase structure; Section 3.2.3 summarizes the temperature and chain length dependencies of unit cell parameters, as determined from the experimental results.

### 3.2.1 Characterization of the studied mixtures

Figures 4, 5, 7 and 8 present DSC and SWAXS spectra for the binary $C_{16} + C_{17}$ and $C_{15} + C_{19}$, ternary $C_{15}+C_{17}+C_{19}$ and the seven-component $C_{14}$-to-$C_{20}$ mixtures. The phase behavior of these samples illustrates all main features observed with the various medium chain length alkane mixtures which is schematically illustrated in Figure 6.

The $C_{16} + C_{17}$ sample is an example of binary mixture in which a complete miscibility is expected, because $\Delta n = n_1 - n_2 = 1 < 4$, as previously shown [45]. In contrast, the chain length difference in $C_{15} + C_{19}$ mixture is $\Delta n = 4$, hence a phase separation is expected, see eq. 1. The behavior of the ternary mixture $C_{15} + C_{17} + C_{19}$ and that of the seven-component equimolar mixture $C_{14}$-to-$C_{20}$ is of interest as well, as these samples represent more intricate systems wherein two distinct chain length differences can be identified: the maximal difference between the shortest and longest alkane in the mixture, $\Delta n_{max} = n_{max} - n_{min} = 4$ or 6, while the minimal chain length difference is $\Delta n_{min} = 2$ or 1. Therefore, these mixtures are examined to determine whether their phase behavior aligns more with mixtures forming solid solutions or with mixtures that phase separate upon liquid-to-solid transition. The results for all studied mixtures are summarized in Table 2 below and in Supporting Information Tables S3 and S4.

## $C_{16}$ + $C_{17}$ mixture ($\Delta n = 1$)

The results for the phase behavior of the simplest binary mixture with $\Delta n = 1$, $C_{16}$ + $C_{17}$, are presented in Figure 4. A liquid-to-rotator phase transition is observed for this sample at $T \approx$ 16°C upon gradual cooling with 0.75°C/min from melt. Note that this temperature is lower than the respective bulk melting points of both alkanes, $T_m(C_{16}) \approx 18°C$ and $T_m(C_{17}) \approx 22°C$. The mixed rotator phase formed is of type $R_I$ with an orthorhombic lattice, which is typical for the odd-numbered alkanes, despite the fact that the studied mixture contains one odd-numbered alkane and one even-numbered alkane. This mixed $R$ phase persists in the sample down to *ca.* -6°C, at which point the $R$-to-$C$ phase transition occurs. The crystallization process (rotator-to-crystal phase transition) within the $C_{16}$ + $C_{17}$ alkane mixture ends at *ca.* -10°C, see Figure 4a.

During heating, the sequence of phase transitions occurred in the reverse order: the crystalline phase transformed into a rotator phase initially, which persisted until the sample reached its final melting point. This observation shows that the $R_I$ phase is thermodynamically stable. The complete melting of the sample ended at $T \approx 19°C$. Notably, this temperature is slightly lower than the bulk melting temperature of pure $C_{17}$, $T_m(C_{17}) \approx 22°C$ [8]. This reduction in melting temperature can be attributed to the molecular solubility of the heptadecane molecules within the shorter hexadecane molecules which possess a lower melting temperature, $T_m(C_{16}) \approx 18°C$. Additionally, the inherent defects within the $R$ phase structure, arising from the chain length mismatch, contribute to a decrease in the overall enthalpy measured upon phase transition in the mixture. The total enthalpy for the mixture is a sum of the enthalpies for $L$-to-$R$ and $R$-to-$C$ transitions, $\Delta H_{tot}$ ($C_{16}+C_{17}$) = $\Delta H_{LR}$ + $\Delta H_{RC} \approx 214.5 \pm 2.5$ J/g. In contrast, the phase transition enthalpies for pure $C_{16}$ and $C_{17}$ alkanes are $\approx 215$ J/g and 230 J/g, respectively, viz. the enthalpy of the 1:1 molar mixture could be as high as 222 J/g. The enthalpy of the $L$-to-$R$ transition in the mixture, $\Delta H_{LR}$, constitutes about 87% of the total enthalpy.

A comparison of the interlamellar spacing observed in the rotator phases of pure hexadecane and heptadecane, and their equimolar mixture is presented in Figure 4d. In the mixed $C_{16}$ + $C_{17}$ rotator phase, the interlamellar spacing is intermediate between that of the two individual alkanes in the mixture. Specifically, it is roughly by 0.5 Å shorter compared to the one measured with $C_{17}$, and by about 0.7 Å longer compared to that with $C_{16}$. The temperature-resolved analysis further reveals that the rate of change in *d*-spacing with temperature for the mixed $C_{16}$ + $C_{17}$ rotator phase is about three times lower compared to that observed in pure $C_{17}$. Despite this, the slope of the area per molecule variation with temperature, $dA/dT$, is only about 20% lower than that for $C_{17}$, although the temperature interval of existence of the *R* phase in $C_{16}$ + $C_{17}$ mixture is about twice bigger compared to that for pure $C_{17}$. Consequently, the mixed $C_{16}$ + $C_{17}$ rotator phase accommodates a 60% larger alteration in the area per molecule before transitioning into *C* phase. This corresponds to $\Delta A \approx 0.5$-$0.6$ Å$^2$ in the mixed system compared to $\Delta A \approx 0.3$-$0.4$ Å$^2$ in the pure alkane systems.

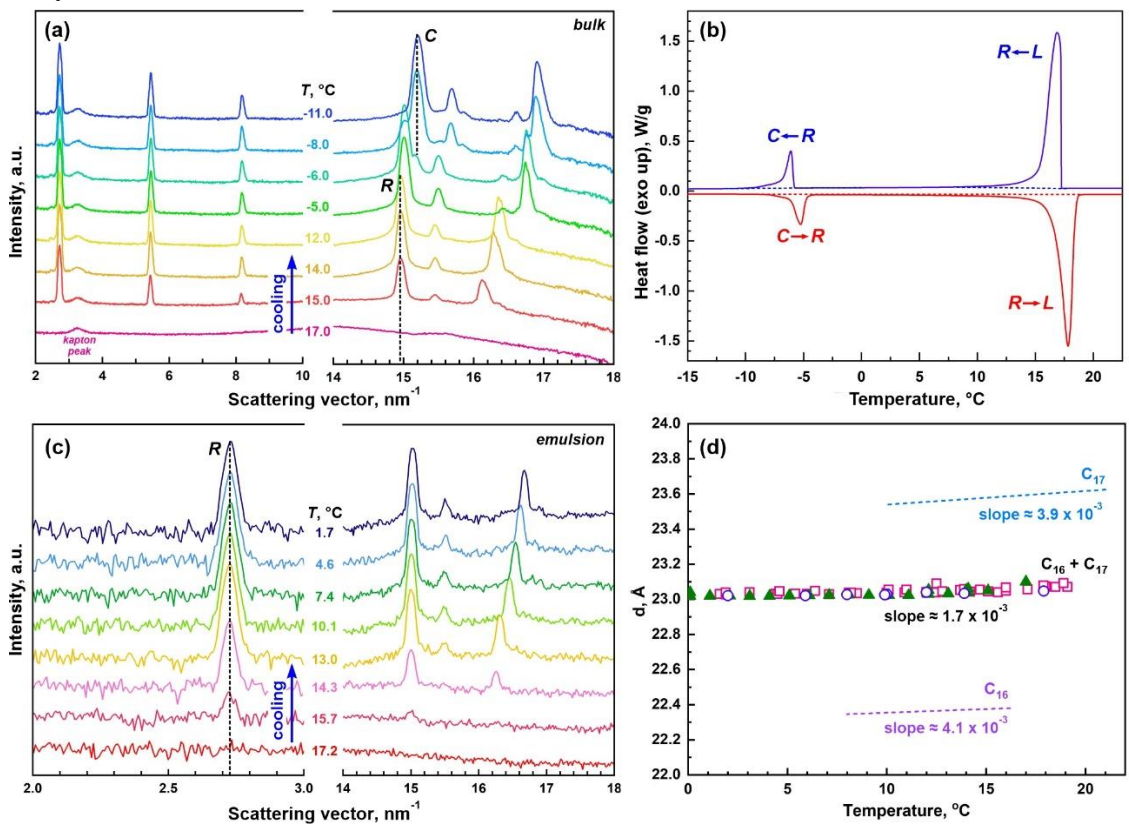

**Figure 4. Phase behavior of $C_{16}$ + $C_{17}$ alkane mixture. (a,c)** SAXS spectra obtained upon cooling of the bulk (a) and emulsified (c) mixture. Note that the *R*-to-*C* phase transition is observed below 0°C, hence it is not seen in the spectra for emulsified sample which is studied at $T > 0$°C to prevent the aqueous phase crystallization. The studied emulsion contains drops with $d \approx 18$ μm, which are stabilized by $C_{16}EO_8$ surfactant. **(b)** DSC thermogram obtained upon cooling and heating of the studied bulk mixture. **(d)** Comparison of the interlamellar spacing of the mixed $C_{16}$ + $C_{17}$ rotator phase determined in three independent experiments: empty pink squares: emulsion sample, synchrotron measurements; empty blue circles: emulsion sample, Xeuss 3.0 experiment, and full green triangles: bulk mixture, Xeuss 3.0. The dashed lines represent the interlamellar spacings for the rotator phases of pure alkanes – blue for $C_{17}$, data from Section 3.1, and purple for $C_{16}$ – data from experiments with emulsion samples presented in Ref. [21].

We note that no significant differences were observed between the experimental results obtained with bulk and emulsified alkane mixtures, see Figure 4a,b,d. Therefore, experiments on emulsions with a synchrotron radiation source were used to characterize the $R$ phase structures within the mixed alkanes. Furthermore, studies were conducted to assess the influence of the initial drop size and of the surfactant molecular characteristics on the rotator phase structure, see Section 3.2.2 below. For some of the examined samples, experiments with bulk alkanes were performed as well, aiming to demonstrate directly the $R$-to-$C$ phase transition. This approach was adopted to eliminate any potential impact from the aqueous phase crystallization on the observed scattering spectra at subzero temperatures.

**$C_{14}$-to-$C_{20}$ mixture ($\Delta n_{min} = 1$, $\Delta n_{max} = 6$)**

The experimental results obtained with the most complex 7-component mixture studied, containing all alkanes with chain lengths between 14 and 20 C-atoms, mixed in equimolar ratio, are presented in Figure 5. The main phase transitions observed in this sample remain qualitatively similar to those observed in the simpler $C_{16} + C_{17}$ mixture, following the $L$-to-$R$-to-$C$ sequence. The $C_{14}$-to-$C_{20}$ mixture comprises alkane homologous with bulk melting temperatures spanning from 5°C (for $C_{14}$) to 37°C (for $C_{20}$). Consequently, the temperature ranges in which distinct phase transitions are observed deviate notably from those observed in the binary $C_{16} + C_{17}$ mixture, see Figure 5d.

The $L$-to-$R$ transition within the $C_{14}$-to-$C_{20}$ mixture, observed upon 0.75°C/min cooling from the melt, begins with a significant subcooling, occurring at a temperature slightly below 20°C. This transition spans a broad temperature interval. The temperature range in which about 78% of the total crystallization enthalpy is released by the alkane mixture is about 16°C (viz. between 19°C and 3°C) for $C_{14}$-to-$C_{20}$ sample, see Figure 5d,e. In contrast, the equivalent range for the $C_{16} + C_{17}$ mixture is only 4°C, see Supporting Information Figure S2 for comparison. Moreover, approximately 57% of the total enthalpy of the $L$-to-$C$ transition is released for $C_{16} + C_{17}$ mixture during the first degree Celsius following the initiation of $L$-to-$R$ transition, in contrast to merely 16% for the $C_{14}$-to-$C_{20}$ mixture. Similar results are obtained when the SAXS peak areas are analyzed as a function of the temperature.

The gradual phase transition within the complex 7-component mixture notably influences the interlamellar spacing, see Figure 5a,e. The initial interlamellar spacing, determined from the bulk experiments (where the nucleation begins at slightly higher temperatures compared to the emulsion experiments) is *ca.* 26 Å. This value diminishes to $d \approx 24.6$ Å at $T \approx 9°C$ where almost all molecules adopt the *R* phase arrangement, see the steep decrease in the *d*-spacing shown in Figure 5e. A similar decrease in $d(T)$ was observed in the emulsion experiments, albeit yielding a slightly shorter final *d*-spacing, most probably reflecting the ordering influence of the surfactant adsorption layer, see the empty pink squares in Figure 5e and the related discussion for the role of surfactant in Section 3.2.2. Upon further temperature decrease, the rate of change in the $d(T)$ function diminishes significantly, decreasing by an order of magnitude (from *ca.* 0.13 Å/°C to 0.013 Å/°C) and becomes comparable to that observed in bulk pure alkanes or in the simpler $C_{16}$ + $C_{17}$ binary mixture where all molecules arrange in *R* phase. Summarized data can be found in Table 2 below.

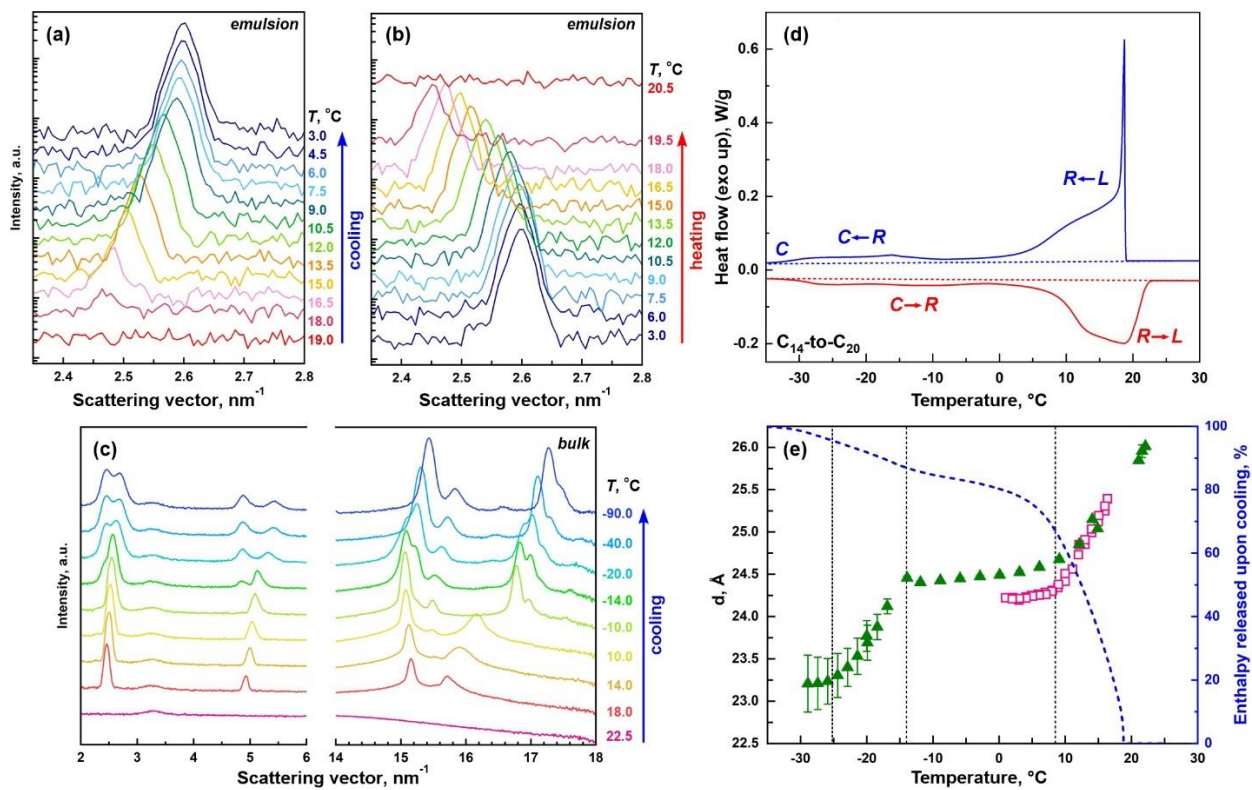

**Figure 5. DSC and SAXS spectra for seven-component equimolar mixture $C_{14}$-to-$C_{20}$. (a,b)** SAXS spectra obtained upon cooling (a) and subsequent heating (b) of 18 µm emulsion droplets, dispersed in $C_{18}EO_{20}$ surfactant. **(c)** SWAXS spectra obtained upon cooling of the bulk mixture. **(d)** DSC thermogram. **(e)** Interlamellar *d*-spacings determined from the bulk (full green triangles) and emulsion experiments (empty pink squares). The dashed blue line shows the cumulative enthalpy function obtained by integration of the cooling DSC thermogram shown in (d). The three vertical black dashed lines denote the regions in which the *d*-spacing changes abruptly its temperature dependence. See text for more detailed explanations.

The 1.4 Å decrease of the $d$-spacing observed in the beginning of the $L$-to-$R$ phase transition (at $T > 10°C$) is slightly bigger than the space required for a single -$CH_2$- group, *i.e.* for the presence of kink defect corresponding to a sequence *gauche-trans-gauche* in an otherwise stretched chain ($\Delta L \approx 1.27$ Å in this case [62]). Notably, although the $R$ phase formation proceeds in a wide temperature interval, all molecules are observed to form a single mixed rotator phase. This result suggests that the individual molecular species remain mixed together during the $L$-to-$R$ phase transition without phase separation into distinct domains with predominant content of longer or shorter alkanes (which would crystallize independently, resulting in the emergence of two or more distinctive interplanar spacings). A plausible molecular arrangement leading to this outcome is depicted schematically in Figure 6a. Note that this behavior is distinctly different from the one observed with mixed monoacid triglyceride systems studied previously [67].

Interestingly, the alterations in the interlamellar spacing in response to temperature changes are completely reversible. Upon heating the samples, an opposite shift in the peak maximum towards lower scattering vectors is observed, see Figure 5b. A comparison between the interlamellar spacing as a function of the temperature in the sample observed upon cooling and heating, shows that the two curves are slightly shifted with respect to the temperature owing to the subcooling phenomenon observed upon cooling. However, the temperature derivatives observed upon cooling and heating are very similar, see Supporting Information Figure S3a.

The rotator-to-crystal phase transition occurs over a wide temperature range, starting at around -10°C and extending to $T \approx -30°C$, see Figure 5c,d. This phase transition led to phase separation of the alkane molecules, resulting in the formation of two phases with distinct interlamellar spacings, $d \approx 23.4$ Å and 25.7 Å (determined at $T \approx -40°C$). These values are comparable to those measured with pure $C_{17}$ and $C_{19}$ at this temperature, $\approx 23.5$ Å and 26.0 Å, respectively. Analysis of peak areas reveals that around 60% of all molecules are arranged within the phase with the shorter $d$-spacing, while the remaining 40% constitute the second phase. However, considering that the $C_{19}$ and $C_{20}$ molecules collectively represent about 28.6% of all alkane molecules in the sample, it becomes evident that the crystal phase characterized by a longer $d$-spacing encompasses molecules with $n < 19$ as well.

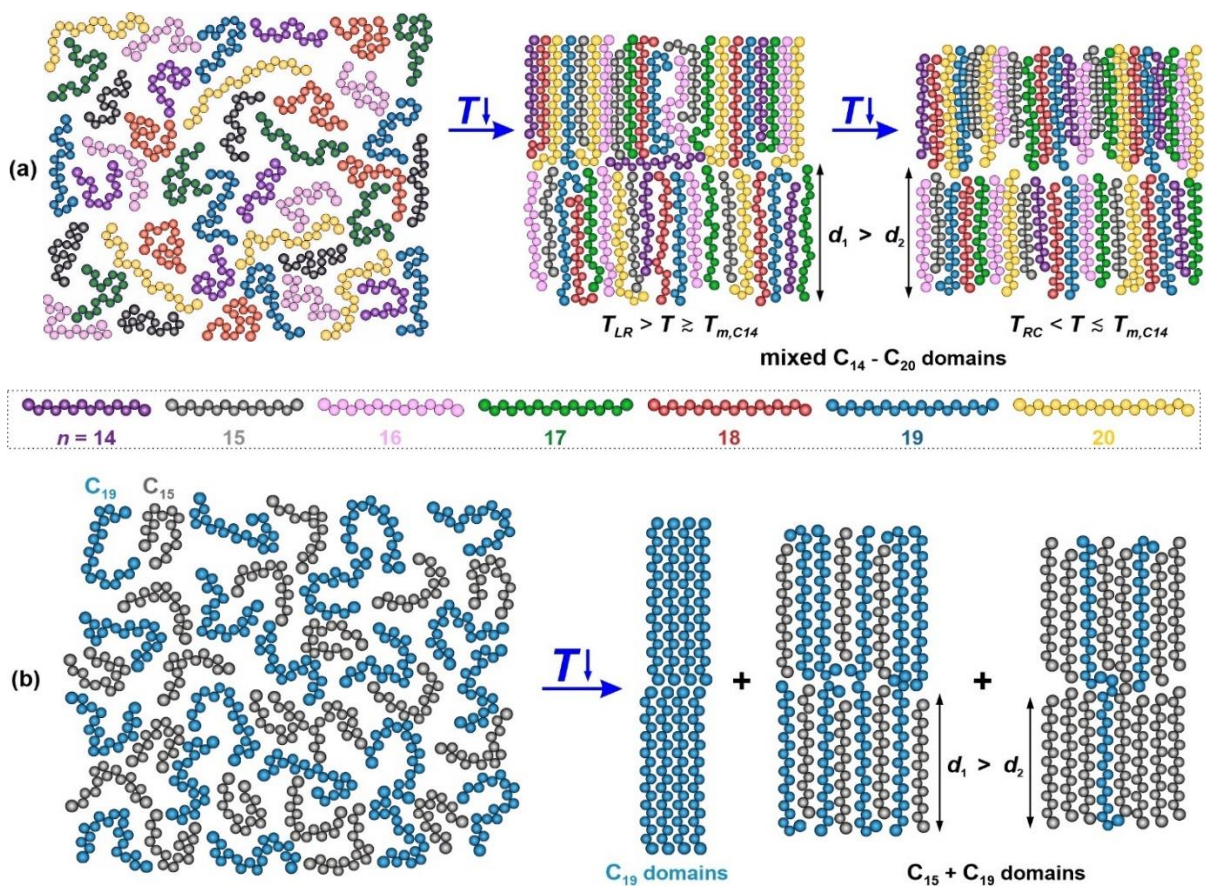

**Figure 6.** **Schematic representation of the mixing/demixing behavior within rotator phases formed in samples of alkanes with various chain lengths.** (**a**) $C_{14}$-to-$C_{20}$ mixture. All molecules arrange together in a single mixed $R$ phase. The interlamellar thickness evolution observed in the temperature range between the onset of $R$ phase formation and the bulk melting temperature of the shortest alkane in the mixture is also schematically illustrated. (**b**) $C_{15} + C_{19}$ binary mixture. Pure $C_{19}$ domains are observed to co-exist with two types of mixed $C_{15} + C_{19}$ domains with different characteristic interlamellar spacing, due to the different $C_{15}$ and $C_{19}$ fraction contained within these two types of domains. Note that in the bulk samples, pure $C_{15}$ domains ($\approx$ 1% peak area) are also observed.

The domain size analysis using Sherrer equation [68] showed similar temperature dependence as the one observed for $d(T)$, see Supporting Information Figure S3b. The average domain size calculated for the emulsion samples at temperatures between 1 and 9°C was $D \approx 140$ nm $\pm$ 16 nm, whereas at higher temperatures the domain size showed slow linear increase with temperature at a rate of $dD/dT \approx 3$ nm/°C.

The obtained results from the $C_{14}$-to-$C_{20}$ mixture reveal that the inclusion of alkanes with intermediate chain lengths between the shortest and the longest components in the mixture suppresses the anticipated phase separation in *R* phase. However, this effect does not prevent the phase separation when the crystalline phase emerges. Accordingly, no phase separation was observed within the rotator phases of the investigated emulsified ternary mixtures, $C_{14} + C_{16} + C_{18}$ and $C_{15} + C_{17} + C_{19}$, see Table 2 below.

**$C_{15} + C_{19}$ mixture ($\Delta n = 4$)**

Next, as an example of a system in which phase separation is observed upon liquid-to-rotator phase transition, we present and discuss the behavior of $C_{15} + C_{19}$ binary equimolar alkane mixture. Upon cooling from the melt, the formation of $C_{19}$ domains was observed, Figure 7a,b and schematics shown in Figure 6b. However, the amount of these domains was significantly lower than the total quantity of $C_{19}$ molecules present in the mixture (50%).

The peak corresponding to pure $C_{19}$ comprised ≈ 3-4% of the total peak area in emulsified samples, and increased to about ≈ 20-25% when investigating bulk $C_{15} + C_{19}$ mixture, Figure 7. The remaining $C_{19}$ molecules formed a mixed rotator phase with the shorter pentadecane molecules. Furthermore, formation of pure $C_{15}$ domains was observed when the bulk samples were studied. The area of this peak represented to a mere 1% of the total peak area, Figure 7d. Although this phase was not identified in the emulsified samples, its existence cannot be entirely ruled out, considering that the intensity of the respective peak might not be sufficient to observe the latter in the background noise (note that the oil volume fraction in the studied emulsions is about 10-20% only).

The rotator-to-crystal phase transition in the bulk sample started at $T \approx -9°C$ and finished at $T \approx -20°C$, see the WAXS curves in Figure 7b and the DSC thermogram in Supporting Information Figure S4. The partial mixing/demixing observed in the rotator phase was preserved also in the crystalline phases, Figure 7b.

Interestingly, within the $C_{15} + C_{19}$ mixture, the formation of (at least) two distinct mixed rotator phases with different interlamellar spacings is observed, see the asymmetric SAXS peaks in Figure 7a,b and the examples for their deconvolution analyses presented in Figure 7c,d. The precise interlamellar spacings determined from these analyses are shown with empty symbols in Figure 7e. These values closely resemble that of pure $C_{17}$ alkane. The dependence $d(T)$ for this mixture follows a similar course to that observed in the $C_{14}$-to-$C_{20}$ mixture. A rapid decrease in the $d$-spacing ($d(c/2)/dT \approx 0.33$ Å/°C) is observed upon the onset of *R* phase, occurring at temperatures above the bulk melting temperature of pure $C_{15}$. As the temperature drops below the bulk melting temperature of $C_{15}$, this rate diminishes.

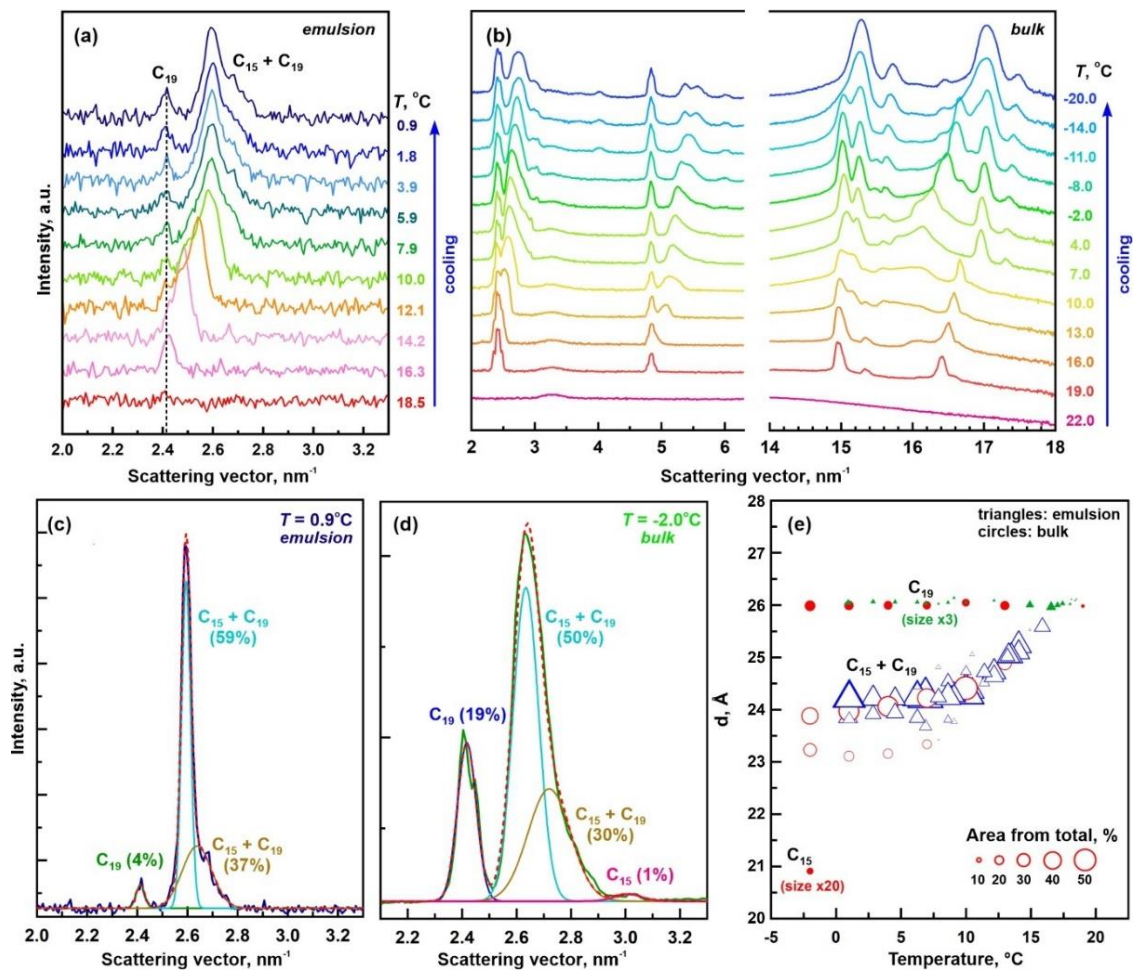

**Figure 7. Structural characterization of the binary $C_{15}$ + $C_{19}$ mixture.** (a,b) SAXS/WAXS spectra obtained upon cooling from melt of emulsified (a) and bulk (b) alkane mixture. The emulsion contains 18 μm drops, stabilized by $C_{16}EO_8$ surfactant. (c) Peak deconvolution analysis of the SAXS spectra obtained at $T = 0.9°C$ for the sample shown in (a). (d) Peak deconvolution analysis of the SAXS spectra obtained at $T = -2.0°C$ for the bulk mixture. (e) Interlamellar spacing as a function of temperature for the bulk (red circles) and emulsified (blue and green triangles) samples. The size of the symbols in the graph corresponds to the peak area of the respective phase. The size of the full green triangles (showing the $d$-spacing in the $C_{19}$ domains) has been increased 3 times to make them visible on the graph, and the size of the red circle showing the formation of $C_{15}$ domains in the bulk sample has been increased 20 times.

We note that similar phase separation behavior was reported for $C_{10}$ + $C_{20}$ binary ($\Delta n = 10$) and the ternary $C_{10}$ + $C_{14}$ + $C_{20}$ ($\Delta n_{min} = 4$ and $\Delta n_{max} = 10$) alkane mixtures in Ref. [48]. Using molecular dynamics simulations, the authors showed that these mixtures crystallized forming domains of single alkane species.

### $C_{15} + C_{17} + C_{19}$ mixture ($\Delta n_{min} = 2$, $\Delta n_{max} = 4$)

Finally, we present results for the phase behavior of the ternary $C_{15} + C_{17} + C_{19}$ mixture, see Figure 8. Our aim is to clarify the impact of introducing a third component, with an intermediate chain length, into the partially phase separating $C_{15} + C_{19}$ mixture.

The SAXS spectra obtained with emulsified mixture, Figure 8a, showed the formation of a single mixed $R$ phase with an interlamellar spacing of $d \approx 24.9$ Å at high temperature ($\approx 17.5$°C). This value decreased to $d \approx 24.0$ Å at $T \approx 2$°C. Once again, two main regions were observed in the $d(T)$ plot – a steep decrease observed at $T > T_m(C_{15})$, followed by a region in which the interlamellar spacing decreased at rate similar to that observed with pure alkanes, see Supporting Information Figure S5a. Similarly, two distinct regions were observed in the domain size dependence on temperature for this mixture, see Supporting Information Figure S5b. In the low temperature range ($T \leq 12$°C), the average crystalline domain size was $D \approx 120$ nm ± 25 nm, whereas at $T > 12$°C – the domain size increased at a rate of $dD/dT \approx 6$ nm/°C upon increase of temperature.

In contrast, the bulk mixture of $C_{15} + C_{17} + C_{19}$ showed a different phase behavior, see Figure 8b. Instead of a single Gaussian peak, as in the case of the emulsified sample, the spectra for the bulk ternary mixture revealed broad and asymmetric peaks for the rotator phase peaks (these are (002) and (004) peaks). This result suggests the presence of various rotator phase domains within the sample. Along with a phase similar to that found in the emulsified sample, three additional domain types were identified, allowing to perform a sufficiently good peak deconvolution analysis, see Supporting Information Figure S5c. These phases exhibited similar $d$-spacings, spanning the range between ca. 23 and 25 Å. This similarity indicates that various molecule types likely coexist within these phases, albeit possibly with different ratios or with various degrees of structural defects within the lamellar structure, see Figure 6b where similar phenomenon in $C_{15} + C_{19}$ mixture is illustrated.

Similar results were obtained also for $C_{15} + C_{17}$ and $C_{17} + C_{19}$ binary alkane mixtures for which $\Delta n = 2$, see Figure 9 for illustrative results with $C_{15} + C_{17}$ mixture and Supporting Information Figure S6 for results obtained with $C_{17} + C_{19}$ mixture. For these samples, we observed the formation of a single mixed rotator phase in the emulsified samples, whereas at least two different types of domains formed in the bulk samples. These results further confirm that the observed phase behavior in the alkane mixtures is primarily governed by the smallest chain-length difference which can be defined for a given mixture, as well as by the confinement size.

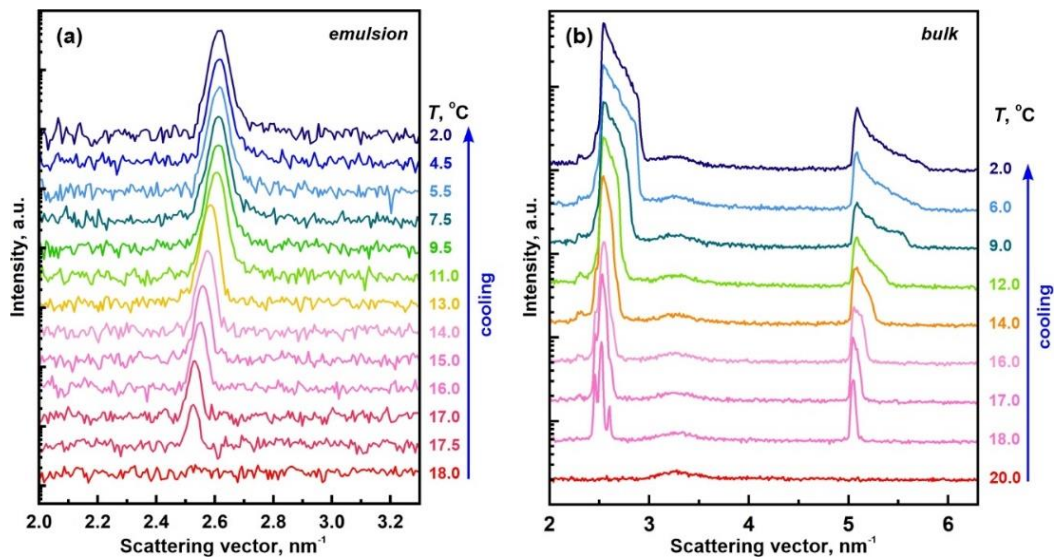

**Figure 8. SAXS spectra of $C_{15}$ + $C_{17}$ + $C_{19}$ ternary mixture. (a)** Emulsion sample, prepared with $C_{18}EO_{20}$ surfactant solution, $d_{ini} \approx 18$ μm. **(b)** Bulk mixture.

In the next Section 3.2.2 we discuss the impact of the initial drop size and of surfactant type on the structure of the mixed rotator phases. An overview of the unit cell parameters and their temperature dependency is provided in Section 3.2.3 and in Supporting Information Table S4.

**3.2.2 Effect of the initial drop size and of surfactant type for the *R* phases formed in mixed alkane droplets**

The effect of the molecular characteristics of the surfactant used for emulsion stabilization (hydrophobic chain length and hydrophilic headgroup size), along with the initial drop size, was systematically explored concerning the structure of the rotator phases formed within the mixed $C_{15}$ + $C_{17}$ emulsified alkane system. This binary mixture has chain length difference $\Delta n = 2$ and, therefore, it forms a single mixed rotator phase. The $d(T)$ dependence is expected to be similar to that observed in $C_{14}$-to-$C_{20}$ or $C_{15}$ + $C_{19}$ mixtures, see Figures 5e and 7e above, *i.e.* a steeper decrease in the interlamellar spacing at temperatures higher than the bulk melting temperature of the shortest alkane in the mixture (in this case, $C_{15}$), followed by a second region in which the interlamellar spacing experiences considerably slower alterations. This behavior was indeed observed experimentally, see Figure 9.

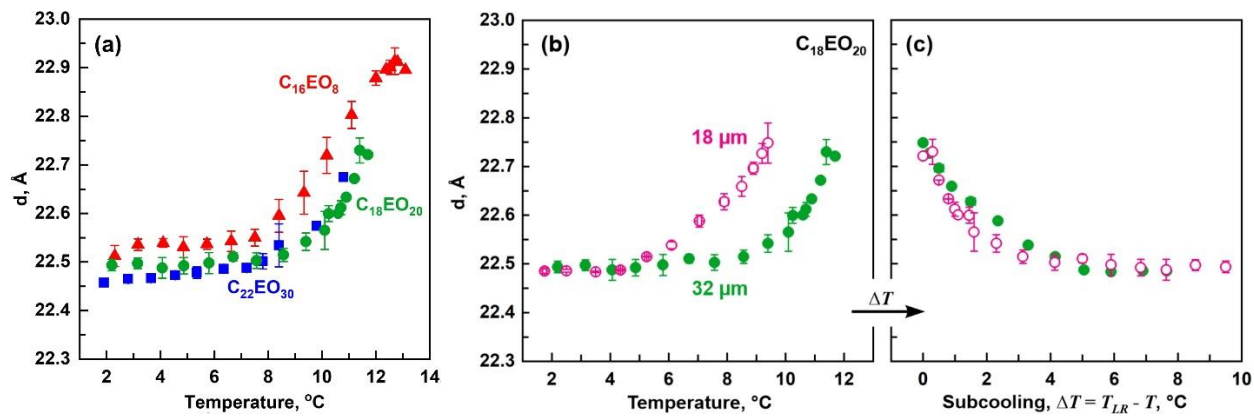

**Figure 9. Effect of surfactant type and initial drop size on the $C_{15} + C_{17}$ mixed rotator phase structure.** (a) Interlamellar spacing measured for 32 μm drops, stabilized by different surfactants: $C_{16}EO_8$ (red triangles), $C_{18}EO_{20}$ (green circles) and $C_{22}EO_{30}$ (blue squares). (b-c) Interlamellar spacing for 18 μm (empty pink circles) and 32 μm (full green circles) drops, stabilized by $C_{18}EO_{20}$ surfactant. The same dataset is presented as a function of temperature (b) and as a function of the subcooling in the sample (c). The subcooling, $\Delta T$, is defined as: $\Delta T = T_{LR} - T$, where $T_{LR}$ is the temperature at which the rotator phase is initially detected in the sample ($T_{LR} \approx 11.7°C$ for 32 μm drops and $\approx 9.4°C$ for 18 μm drops) and $T$ is the temperature at which a given data point was acquired.

Figure 9a presents the experimental results obtained with $C_{15} + C_{17}$ emulsion drops with similar diameters, $d_{ini} \approx 32$ μm, stabilized by non-ionic alcohol ethoxylated surfactants with different structures, $C_nEO_m$. The hydrophobic tail length, $n$, was varied between 16 and 22 C-atoms, while the number of sequentially connected ethoxy units, $m$, ranged from 8 to 30. Previous studies involving similar emulsion systems demonstrated that, upon cooling, the drops spontaneously undergo shape transformations, transitioning into diverse fluid polyhedra, triangular, tetragonal or hexagonal platelets, rod-like particles or even thin fibers [69]. The currently studied emulsions also undergo drop shape transformations prior to the formation of rotator phase in the entire drop volume, see Supporting Information Figure S7. These deformations are governed by the formation of thin multilayers of plastic rotator phases, adjacent to the drop surface. These ordered layers (with thickness between ca. 2-3 up to 20-30 layers) appear before the entire oily phase contained within the droplets transition from a liquid state to a rotator phase [21].

The newly obtained results show that the surfactant structure affects to some extend the *d*-spacing observed at low temperatures. Surfactants with longer tails prompt the formation of phases with shorter interlamellar spacing, whereas slightly longer repeat distance of the layers is observed at equivalent temperatures when the same drops are dispersed in surfactant with shorter hydrophobic tail, Figure 9a. The average difference in the thickness between the rotator phases formed in drops, stabilized by $C_{16}EO_8$ and $C_{22}EO_{30}$ surfactants is approximately 6.5 pm (this comparison is based on data obtained at temperatures between 1 and 5°C). Although this difference is relatively small, the experimental results obtained with synchrotron radiation source were with excellent reproducibility and confirmed this conclusion. Statistical analysis of the averaged data, obtained from 2-3 independent experiments for each emulsion sample proved that the obtained values are statistically different from one another. Furthermore, the structuring role of the surfactant adsorption layer was further confirmed by the results obtained with $C_{18}EO_{20}$ surfactant for which intermediate *d*-values were observed between those for $C_{16}EO_8$ and $C_{22}EO_{30}$ surfactants, see the green circles in Figure 9a.

The results highlighting the effect of the initial drop size are presented in Figure 9b,c. Two distinct samples were investigated, with drop diameters of ≈ 18 and 32 μm. The obtained results for the interlamellar spacings plotted as a function of temperature, Figure 9b, showed a significant difference between these two samples. For example, at *T* = 8.5°C the *d*-spacing is ≈ 22.66 Å in the sample with smaller drops and ≈ 22.51 Å in the sample with bigger drops. Once the temperature is decreased below 5°C, however, this difference diminishes and the interlamellar spacings in the two samples become identical.

Further analysis of the obtained results, showed that the observed effect is governed by the initial temperature at which the rotator phase begins to form. The nucleation process occurs with an approximately 2°C difference between the two samples. This variation is attributed to the decreased probability for nucleation in smaller droplets, a phenomenon described by the well-known subcooling effect. In general, smaller drops require higher degrees of subcooling to trigger the nucleation process, due to their reduced material content [8,70,71]. Therefore, when the same dataset is plotted as a function of the subcooling, $\Delta T = T_{LR} - T$, the two dependences merge together, see Figure 9c. The subcooling value $\Delta T$ is defined as the difference between the temperature at which the rotator phase is initially detected in the sample, $T_{LR}$, and the temperature which the particular data point is collected. This result shows that the confinement size, at least at the micrometer scale, influences the initial nucleation temperature, but it does not affect the rotator phase structure. Note that the present results cannot be directly related to the surface freezing phenomenon and its temperature dependence, as our previous results with similar oil-in-water emulsions showed that the SAXS experiments, even when performed with synchrotron radiation source, are unable to detect the formation of 1-2 layers of ordered molecules [21].

### 3.2.3 Discussion of the temperature dependence of the unit cell parameters

Table 2 summarizes the results obtained with various emulsion samples studied. These results show that the phenomenon of phase separation within these samples is determined by the smallest chain length difference in the mixture, $\Delta n_{min}$. For example, $C_{14}$-to-$C_{20}$ mixture arranges into a single rotator phase, whereas a partial phase separation is observed in the $C_{15}$ + $C_{19}$ mixture. Notably, the predominant fraction of the molecules in $C_{15}$ + $C_{19}$ mixture arrange in a mixed $R$ phase. About 25% of the molecules (mainly $C_{19}$) were observed to phase separate in the bulk sample, whereas this percentage was significantly lower (*ca.* 3-4%) in the emulsified samples. The addition of hydrocarbons with intermediate chain length suppresses the phase separation within the rotator phase. These findings are schematically illustrated in Figure 6. The obtained data evidence that the temperature range of existence of the mixed $R$ phase expands when larger number of components with larger disparities in their chain lengths are mixed.

The interlamellar spacings for the $R_I$ phases determined in the current study are in general agreement with the spacings that can be calculated using the equation derived by Dirand and co-authors, see eq. 2 above [7,49]. The newly obtained data show that the deviation between the calculated interlamellar spacing and the one measured experimentally is linearly proportional to the greatest chain length difference within the mixture, see Supporting Information Figure S8. The slope of this dependence equals 0.1268. Consequently, we propose a refined equation that accounts for the increased lamellar surface roughness and the presence of void volumes:

$$d = \frac{c}{2} \approx 1.2724(\bar{n}-1) + 3.1476 + 0.1268\Delta n_{max} \text{ Å} \qquad (6)$$

Using this equation, we calculated successfully the observed interlamellar spacings in the mixed $R$ phases at $T \approx 2°C$ ($R^2 = 0.995$) for all studied mixtures, except for $C_{14}$ + $C_{16}$ + $C_{18}$. This particular mixture comprised even-numbered alkanes, which are known to exhibit very short living transient $R$ phases within their bulk samples and metastable $R$ phases within emulsion drops when taken as individual components [9,19-21]. Instead, the mixing of these alkanes results in the formation of a thermodynamically stable rotator phase in a wide temperature interval. However, the interlamellar spacing of the *R* phase measured in the absence of $C_{14}$ (for $C_{16}$ + $C_{18}$ system) or in its presence within the ternary mixture, remained practically the same. This observation shows that the inclusion of $C_{14}$ molecules does not influence the interlamellar spacing. The root cause of this deviation from the general trend requires further in-depth investigation. Such effect was absent when analyzing mixtures of odd-numbered alkanes, such as $C_{15}$ + $C_{17}$ and $C_{15}$ + $C_{17}$ + $C_{19}$, where the addition of $C_{15}$ led to an overall decrease in the lamellar spacing, as expected, due to the decrease in the molar average chain length within the sample.

Similarly to the rotator phases formed by pure alkanes, the unit cell parameter *a* in the mixed *R* phases decreased as the temperature decreased, see Supporting Information Table S4. Conversely, the shorter length of the orthorhombic unit cell, parameter *b*, increased as the temperature decreased. The *a(T)* and *b(T)* dependencies showed linear behavior at temperatures slightly below the onset of rotator phase formation. The slopes of these trends depended on the complexity of the mixture – mixtures with larger $\Delta n_{max}$ exhibited greater slopes (in absolute values). Nonetheless, the temperature resolved structural analysis of the mixtures showed a consistent decrease in the area per molecule at a rate of $(2.7 \pm 0.3) \times 10^{-2}$ Å$^2$/°C across all investigated mixtures with molar average chain lengths $\bar{n}$ between 16 and 17. Notably, this value is identical to the one obtained for pure heptadecane, see Figure 3a and Supporting Information Table S1.

As elucidated in the previous two sections, the *c(T)* dependencies for mixed samples with $\Delta n > 1$ exhibit two distinct regions. At elevated temperatures exceeding the bulk melting point of the shortest alkane in the mixture, *c(T)* function experiences a rapid decline. This decline is attributed to the ongoing *L*-to-*R* phase transition proceeding in the sample. As the temperature within the sample approaches the melting point of the shortest alkane in the mixture, the rate of change in the *c(T)* dependence decreases by about 10 times, approaching the values observed for *R* phases of pure alkanes.

The thermal expansion coefficients (see eq. 5) for rotator phases of mixed alkanes are summarized in Supporting Information Table S3. Notably, the volumetric expansion coefficient for rotator phases of mixed alkanes turn out to be about 10 times higher compared to those observed for pure alkanes, $\alpha_{V,R} \approx 2.0 \times 10^{-3}$ °C$^{-1}$ for alkane mixtures, whereas $\alpha_{V,R} \approx 1.5 \times 10^{-4}$ °C$^{-1}$ for the rotator phases of $C_{13}$ to $C_{19}$ bulk pure alkanes.

**Table 2.** Summarized data obtained from the experiments with alkane mixtures. $T_{LR}$ is the temperature at which the rotator phase is initially detected, and $\Delta T_R = T_{RL} - T_{RC}$ is the temperature range in which it is observed to exist before the transition into the crystalline phase begins. These temperatures are determined upon cooling from SWAXS cooling experiment with bulk samples. Their accuracy is about ± 2°C, as in most of the experiments we took a single spectrum at each 2°C. The rest of the data presented in the table is obtained from experiments with emulsion samples ($T > 0°C$). Similar data for the pure alkanes is available in Supporting Information Table S1.

| Alkane mixture | $\Delta n_{max}$ | $\Delta n_{min}$ | $\bar{n}$ | $T_{LR}$, °C | $T_{RC}$, °C | $\Delta T_R$, °C | Phases formed upon cooling (emulsions) | $d$ at $T = 2°C$, Å | $d(c/2)/dT$, Å/°C High T stdev ≈ 0.02 | $d(c/2)/dT$, Å/°C Low T | $da/dT$, Å/°C | $db/dT$, Å/°C | $dA/dT$, Å²/°C |
|---|---|---|---|---|---|---|---|---|---|---|---|---|---|
| $C_{16} + C_{17}$ | 1 | 1 | 16.5 | 16 | -6 | 22 | 1 mixed | 23.0 | - | 0.002 | 0.015 | -0.004 | 0.023 |
| $C_{15} + C_{17}$ | 2 | 2 | 16 | 12 | -20 | 32 | 1 mixed | 22.5 | 0.10 | 0.008 | 0.021 | -0.007 | 0.027 |
| $C_{16} + C_{18}$ | 2 | 2 | 17 | 20 | -10 | 30 | 1 mixed | 23.8 | 0.08 | 0.003 | 0.023 | -0.007 | 0.032 |
| $C_{17} + C_{19}$ [1] | 2 | 2 | 18 | 24 | -3 | 27 | 1 mixed | 24.7 | 0.05 | 0.002 | 0.022 | -0.007 | 0.025 |
| $C_{14} + C_{16} + C_{18}$ | 4 | 2 | 16 | 19 | -8 | 27 | 1 mixed | 23.8 | 0.04 | 0.003 | 0.018 | -0.006 | 0.026 |
| $C_{15} + C_{17} + C_{19}$ | 4 | 2 | 17 | 20 | -10 | 30 | 1 mixed | 24.0 | 0.13 | 0.006 | 0.028 | -0.011 | 0.025 |
| $C_{15} + C_{19}$ [2] | 4 | 4 | 17 | 19 | -8 | 27 | 2 mixed R + $C_{19}$ | 23.8 24.1 26.1 ($C_{19}$) | 0.33 | 0.027 | 0.029 | -0.012 | 0.027 |
| $C_{14}$-to-$C_{20}$ [3] | 6 | 1 | 17 | 22 | -12 | 36 | 1 mixed | 24.2 | 0.13 | 0.013 | 0.035 | -0.015 | 0.028 |

(1) Temperature derivatives for $a(T)$ and $b(T)$ are determined for $10°C \leq T \leq 20°C$.

(2) All temperature derivatives are calculated for the mixed $R$ phases.

(3) All temperature derivatives are calculated from data obtained at $T < 12°C$.

## 4. Conclusions

The present study investigates the temperature dependent characteristics of rotator phases formed in both pure and mixed linear alkanes with chain length varied between 13 and 21 C-atoms. Empirical equations have been defined to describe the unit cell parameters for the rotator and crystalline phases of odd-numbered alkanes, as a function of alkane chain length and temperature.

The results obtained from experiments conducted with both bulk and emulsified mixed alkane samples revealed that confinement at the micrometer scale does not influence the structure of the rotator phases. Single mixed $R$ phase was formed in all emulsified samples when the smallest chain length difference between the homologous in the mixture was $\Delta n_{min} \leq 2$. A partial phase separation was observed for the sample with $\Delta n_{min} = 4$. However, even in this sample, the main fraction of the molecules assembled into mixed solid solution. This partial phase separation can be suppressed by introducing an additional component with an intermediate chain length into the sample, thereby reducing $\Delta n_{min}$. Phase separation was not observed in the emulsified samples prepared using such ternary mixtures of alkanes, whereas several diverse domain types with similar interlamellar spacings were detected in the corresponding bulk mixtures.

The interlamellar spacing within mixed rotator phases is primarily determined by the molar averaged chain length in the mixture, as previously demonstrated [7,49]. Additionally, an extra term is introduced to the established relation to account for the increased void volume resulting from the disparity in chain lengths. The present results show that for mixtures which do not phase separate upon cooling and which contain at least one odd-numbered alkane, this additional term is $\approx 0.127\ \Delta n_{max}$.

Interestingly, a significant shift in the SAXS peak position toward higher $q$-values, corresponding to smaller $d$-spacings, was consistently observed across all mixed samples investigated where $\Delta n \geq 1$ at temperatures close to the temperature at which the liquid-to-rotator phase transition began. The DSC data, along with the scattering peak area analysis, showed that this $L$-to-$R$ phase transition is not instantaneous, rather it occurs over a relatively broad temperature range. Typically, this range extends until the bulk melting temperature of the shortest alkane in the mixture is nearly approached. The observed temperature shift of the peaks at lower temperatures was comparable to that exhibited by the rotator phases of pure alkanes.

The experiments involving emulsions with various drop sizes and diverse surfactants showed that the confinement size primarily impacts the initial temperature at which the formation of the *R* phase starts. Smaller emulsion drops exhibited a more pronounced subcooling, attributed to the reduced probability of rotator phase nucleus formation. The surfactant adsorption layer was found to exert a minor impact on the interlamellar spacing of the *R* phases – slightly shorter *d*-spacings were observed for drops stabilized by surfactants with longer saturated hydrocarbon tails.

These results contribute to a deeper understanding of the intricate structure of intermediate rotator phases. These phases are important in wide array of applications, ranging from engineering, food science, cosmetics, and even essential processes within living organisms.

**Supporting Information:** Table with coefficients describing the linear temperature dependences of lattice parameters for bulk odd-numbered alkanes; Tables with unit cell parameters measured at different temperatures and thermal expansion coefficients for rotator and crystalline phases of the studied alkanes/alkane mixtures; SAXS/WAXS spectra of $C_{13}$ and $C_{17} + C_{19}$ binary mixture; Cumulative enthalpy released upon cooling of $C_{16} + C_{17}$ and $C_{14}$-to-$C_{20}$ alkane mixtures; Comparison – interlamellar spacing measured upon cooling and heating for $C_{14}$-to-$C_{20}$ alkane mixture; Average crystalline domain size as a function of temperature for $C_{14}$-to-$C_{20}$ and $C_{15}+C_{17}+C_{19}$ emulsified alkane mixtures; DSC thermogram for $C_{15} + C_{19}$ mixture; Interlamellar spacing and peak deconvolution analysis for $C_{15} + C_{17} + C_{19}$ alkane mixture; Microscopy pictures obtained upon cooling of $C_{15} + C_{17}$ emulsion droplets; Rotator phase interlamellar spacing dependence on $\Delta n_{max}$.

**Acknowledgements:**

The study was funded by the Bulgarian Ministry of Education and Science, under the National Research Program "VIHREN", project ROTA-Active (no. KP-06-DV-4/16.12.2019). The research leading to part of the results has been supported by the project CALIPSOplus under Grant Agreement 730872 from the EU Framework Programme for Research and Innovation HORIZON 2020, under the execution of project 20217121 with proposer D.C. The authors thank Dr. Barbara Sartori and Dr. Heinz Amenitsch (Elettra Sincrotron, Trieste, Italy) for their valuable help during the SAXS/WAXS synchrotron measurements. The authors are grateful to Ms. Desislava Glushkova and Mrs. Sonya Tsibranska-Gyoreva (Sofia University) for their help during the synchrotron measurements. The authors acknowledge the possibility to use Xeuss 3.0 equipment purchased for execution of project BG05M2OP001-1.002-0012, Operational Program "Science and Education for Smart Growth", Bulgaria.

**Contributions:**

D.C. – conceptualization; methodology; investigation; validation; formal analysis; visualization; writing – original draft, review and editing; project administration; funding acquisition.

M.P. – investigation; visualization; formal analysis; writing – original draft.

S.Tc. – conceptualization; supervision; writing – review and editing; funding acquisition.

N.D. – conceptualization; supervision; writing – review and editing; funding acquisition.

# Supporting Information

## Structure of rotator phases formed in $C_{13}$-$C_{21}$ alkanes and their mixtures: in bulk and in emulsion drops

**Diana Cholakova\*, Martin Pantov**
**Slavka Tcholakova, Nikolai Denkov**

*Department of Chemical and Pharmaceutical Engineering*
*Faculty of Chemistry and Pharmacy, Sofia University,*
*1 James Bourchier Avenue, 1164 Sofia, Bulgaria*

\*Corresponding authors:
Dr. Diana Cholakova
Department of Chemical and Pharmaceutical Engineering
Sofia University
1 James Bourchier Ave.,
Sofia 1164
Bulgaria
E-mail: dc@dce.uni-sofia.bg
Tel: +359 2 8161624

**Supporting Table S1. Linear dependences describing the thermal behavior of rotator and crystalline phases of bulk pure alkanes.** Coefficients for the linear equations of lattice parameters $a$, $b$ and $c$, and the area per molecule $A$. In the last column – the maximal ($A_{max}$) and minimal areas ($A_{min}$) per molecule determined from the experimental data are presented for $R$. For $C$ phase, $A_{max}$ is presented only, as $A_{min}$ depends on the minimal temperature reached in a given experiment. Note that the slopes of the different quantities obtained in $C$ phases are one order of magnitude smaller compared to those for $R$ phases.

| Alkane | $a(T) = xT + y$ | | $b(T) = xT + y$ | | $c/2(T) = xT + y$ | | $A(T) = xT + y$ | | Areas, Å² | |
|---|---|---|---|---|---|---|---|---|---|---|
| | $x$, Å/°C | $y$, Å | $x$, Å/°C | $y$, Å | $x$, Å/°C | $y$, Å | $x$, Å²/°C | $y$, Å² | | |
| *Rotator phases* | | | | | | | | | $A_{max}$ | $A_{min}$ |
| $C_{13}$ | $0.9 \times 10^{-2}$ | 7.69 | $-1.1 \times 10^{-3}$ | 5.08 | $6.0 \times 10^{-4}$ | 18.44 | $2.0 \times 10^{-2}$ | 19.5 | 19.4 | 19.0 |
| $C_{15}$ | $1.3 \times 10^{-2}$ | 7.56 | $-2.3 \times 10^{-3}$ | 5.07 | $5.3 \times 10^{-3}$ | 20.97 | $2.4 \times 10^{-2}$ | 19.2 | 19.4 | 19.1 |
| $C_{17}$ | $1.7 \times 10^{-2}$ | 7.40 | $-3.8 \times 10^{-3}$ | 5.12 | $3.9 \times 10^{-3}$ | 23.57 | $2.7 \times 10^{-2}$ | 18.9 | 19.6 | 19.2 |
| $C_{19}$ | $1.9 \times 10^{-2}$ | 7.22 | $-4.9 \times 10^{-3}$ | 5.16 | $1.4 \times 10^{-2}$ | 25.88 | $2.9 \times 10^{-2}$ | 18.7 | 19.7 | 19.3 |
| $C_{21}$ | $3.9 \times 10^{-2}$ | 6.43 | $-2.1 \times 10^{-2}$ | 5.74 | $5.7 \times 10^{-2}$ | 27.02 | $9.5 \times 10^{-2}$ | 17.8 | 19.8 | 19.5 |
| *Crystalline phases* | | | | | | | | | $A_{max}$ | |
| $C_{13}$ | $1.6 \times 10^{-3}$ | 7.43 | $5.8 \times 10^{-4}$ | 5.02 | $< 1 \times 10^{-4}$ | 18.52 | $6.1 \times 10^{-3}$ | 18.6 | 18.5 | |
| $C_{15}$ | $1.9 \times 10^{-3}$ | 7.42 | $6.7 \times 10^{-4}$ | 4.98 | $< 1 \times 10^{-4}$ | 20.67 | $7.2 \times 10^{-3}$ | 18.5 | 18.6 | |
| $C_{17}$ | $2.3 \times 10^{-3}$ | 7.42 | $5.6 \times 10^{-4}$ | 4.99 | $2.2 \times 10^{-4}$ | 23.50 | $7.8 \times 10^{-3}$ | 18.5 | 18.6 | |
| $C_{19}$ | $2.7 \times 10^{-3}$ | 7.41 | $4.7 \times 10^{-4}$ | 4.98 | $9.0 \times 10^{-4}$ | 26.08 | $8.4 \times 10^{-3}$ | 18.5 | 18.7 | |
| $C_{21}$ | $2.5 \times 10^{-3}$ | 7.42 | $4.1 \times 10^{-4}$ | 4.99 | $1.1 \times 10^{-3}$ | 28.51 | $7.7 \times 10^{-3}$ | 18.5 | 18.7 | |

**Supporting Table S2. Unit cell parameters for bulk odd-numbered pure alkanes measured for different temperatures.** The values calculated using eq. 3 from the main text are also shown for comparison. The values for rotator phases are shown with red, while those for crystalline phases are shown with blue color.

| $T$, °C | Experimentally measured values | | | Calculated values using eq. 3 and coefficients from Table 1 | | |
|---|---|---|---|---|---|---|
| | $a$, Å | $b$, Å | $c/2$, Å | $a$, Å | $b$, Å | $c/2$, Å |
| Tridecane - $C_{13}H_{28}$ | | | | | | |
| -8  | 7.618 | 5.091 | 18.514 | 7.618 | 5.075 | 18.478 |
| -11 | 7.586 | 5.095 | 18.516 | 7.591 | 5.078 | 18.478 |
| -15 | 7.545 | 5.101 | 18.517 | 7.552 | 5.082 | 18.477 |
| -19 | 7.512 | 5.103 | 18.515 | 7.514 | 5.087 | 18.477 |
| -21 | 7.497 | 5.104 | 18.514 | 7.504 | 5.088 | 18.476 |
| -30 | 7.383 | 5.003 | 18.430 | 7.371 | 4.989 | 18.377 |
| -50 | 7.342 | 4.988 | 18.409 | 7.337 | 4.976 | 18.378 |
| -60 | 7.325 | 4.982 | 18.409 | 7.320 | 4.970 | 18.379 |
| -80 | 7.296 | 4.972 | 18.396 | 7.286 | 4.957 | 18.380 |
| -90 | 7.268 | 4.962 | 18.386 | 7.269 | 4.950 | 18.380 |
| Pentadecane - $C_{15}H_{32}$ | | | | | | |
| 10  | 7.683 | 5.050 | 21.027 | 7.669 | 5.071 | 21.011 |
| 8   | 7.672 | 5.053 | 21.014 | 7.643 | 5.076 | 21.003 |
| 4   | 7.611 | 5.063 | 20.979 | 7.592 | 5.085 | 20.987 |
| 0   | 7.559 | 5.072 | 20.977 | 7.540 | 5.094 | 20.971 |
| -5  | 7.407 | 4.986 | 20.897 | 7.411 | 4.997 | 20.915 |
| -14 | 7.388 | 4.978 | 20.640 | 7.393 | 4.991 | 20.913 |
| -24 | 7.369 | 4.972 | 20.462 | 7.373 | 4.985 | 20.910 |
| -60 | 7.298 | 4.948 | 20.458 | 7.303 | 4.964 | 20.903 |
| Heptadecane - $C_{17}H_{32}$ | | | | | | |
| 20  | 7.733 | 5.041 | 23.651 | 7.706 | 5.050 | 23.617 |
| 17  | 7.679 | 5.053 | 23.636 | 7.657 | 5.061 | 23.594 |
| 13  | 7.613 | 5.068 | 23.623 | 7.593 | 5.075 | 23.563 |
| 11  | 7.586 | 5.075 | 23.616 | 7.561 | 5.083 | 23.548 |
| 6   | 7.436 | 4.993 | 23.499 | 7.431 | 4.994 | 23.459 |
| -1  | 7.418 | 4.989 | 23.497 | 7.416 | 4.990 | 23.455 |
| -8  | 7.402 | 4.985 | 23.496 | 7.401 | 4.987 | 23.452 |
| -14 | 7.389 | 4.982 | 23.494 | 7.387 | 4.983 | 23.449 |
| -20 | 7.376 | 4.978 | 23.493 | 7.374 | 4.980 | 23.446 |

**Supporting Table S2 – continued**

| T, °C | Experimentally measured values | | | Calculated values using eq. 3 and coefficients from Table 1 | | |
|---|---|---|---|---|---|---|
| | a, Å | b, Å | c/2, Å | a, Å | b, Å | c/2, Å |
| Nonadecane - $C_{19}H_{40}$ | | | | | | |
| 30 | 7.796 | 5.013 | 26.294 | 7.807 | 5.003 | 26.298 |
| 27 | 7.730 | 5.033 | 26.241 | 7.749 | 5.018 | 26.264 |
| 24 | 7.670 | 5.047 | 26.225 | 7.691 | 5.033 | 26.229 |
| 22 | 7.646 | 5.052 | 26.175 | 7.652 | 5.043 | 26.206 |
| 16 | 7.454 | 4.986 | 26.094 | 7.455 | 4.990 | 26.008 |
| 10 | 7.439 | 4.983 | 26.089 | 7.441 | 4.987 | 26.003 |
| 0 | 7.412 | 4.978 | 26.080 | 7.416 | 4.982 | 25.996 |
| -5 | 7.398 | 4.976 | 26.075 | 7.404 | 4.980 | 25.992 |
| Heneicosane - $C_{21}H_{44}$ | | | | | | |
| 40 | 8.146 | 4.873 | 29.274 | n/a (equation is not applicable) | | |
| 35 | 7.779 | 5.021 | 28.740 | | | |
| 32 | 7.717 | 5.040 | 28.544 | | | |
| 25 | 7.482 | 4.990 | 28.542 | 7.482 | 4.984 | 28.561 |
| 21 | 7.471 | 4.988 | 28.532 | 7.471 | 4.983 | 28.557 |
| 10 | 7.440 | 4.984 | 28.524 | 7.441 | 4.978 | 28.546 |
| 0 | 7.417 | 4.980 | 28.509 | 7.414 | 4.974 | 28.536 |
| -20 | 7.369 | 4.972 | 28.493 | 7.360 | 4.965 | 28.515 |

**Supporting Table S3. Thermal expansion coefficients.** The thermal expansion coefficients are defined as: $\alpha_f = \frac{1}{f_0}\left(\frac{df}{dT}\right)_p$, where $f$ is either of the unit cell parameters ($a$, $b$, or $c$) or the unit cell volume, $V = abc$ for orthorhombic lattice. $f_0$ is the reference value of the studied parameter. For all alkane mixtures – the values measured at temperature $T = 2°C$ were taken as reference values. For the rotator phases of pure alkanes – $f_0$ values were taken at the lowest temperature at which the rotator phase existence is observed, whereas for the crystalline phases – the values measured/calculated for $T = -40°C$ were used as reference values.

| Alkane/alkane mixture | $\alpha_a$, °C$^{-1}$ | $\alpha_b$, °C$^{-1}$ | $\alpha_c$, °C$^{-1}$ | $\alpha_V$, °C$^{-1}$ |
|---|---|---|---|---|
| **Bulk odd-numbered alkanes – rotator phases** | | | | |
| $C_{13}$ | $1.25 \times 10^{-3}$ | $-0.22 \times 10^{-3}$ | $0.03 \times 10^{-3}$ | $1.26 \times 10^{-4}$ |
| $C_{15}$ | $1.73 \times 10^{-3}$ | $-0.45 \times 10^{-3}$ | $0.25 \times 10^{-3}$ | $1.53 \times 10^{-4}$ |
| $C_{17}$ | $2.17 \times 10^{-3}$ | $-0.74 \times 10^{-3}$ | $0.16 \times 10^{-3}$ | $1.47 \times 10^{-4}$ |
| $C_{19}$ | $2.49 \times 10^{-3}$ | $-0.96 \times 10^{-3}$ | $0.52 \times 10^{-3}$ | $1.55 \times 10^{-4}$ |
| $C_{21}$ | $7.26 \times 10^{-3}$ | $-4.14 \times 10^{-3}$ | $3.25 \times 10^{-3}$ | $2.43 \times 10^{-4}$ |
| **Bulk odd-numbered alkanes – crystalline phases** | | | | |
| $C_{13}$ | $2.20 \times 10^{-4}$ | $1.16 \times 10^{-4}$ | $< 1 \times 10^{-5}$ | $5.32 \times 10^{-5}$ |
| $C_{15}$ | $2.63 \times 10^{-4}$ | $1.35 \times 10^{-4}$ | $< 1 \times 10^{-5}$ | $6.30 \times 10^{-5}$ |
| $C_{17}$ | $3.12 \times 10^{-4}$ | $1.13 \times 10^{-4}$ | $< 1 \times 10^{-5}$ | $6.57 \times 10^{-5}$ |
| $C_{19}$ | $3.64 \times 10^{-4}$ | $9.45 \times 10^{-5}$ | $3.5 \times 10^{-5}$ | $7.02 \times 10^{-5}$ |
| $C_{21}$ | $3.43 \times 10^{-4}$ | $8.16 \times 10^{-5}$ | $3.7 \times 10^{-5}$ | $6.46 \times 10^{-5}$ |
| **Alkane mixtures – rotator phases**[1] | | | | |
| $C_{16} + C_{17}$ | $2.06 \times 10^{-3}$ | $-0.77 \times 10^{-3}$ | $- / 0.11 \times 10^{-3}$ | $1.37 \times 10^{-3}$ |
| $C_{15} + C_{17}$ | $2.75 \times 10^{-3}$ | $-1.39 \times 10^{-3}$ | $4.44 \times 10^{-3} / 0.36 \times 10^{-3}$ | $2.32 \times 10^{-3}$ |
| $C_{16} + C_{18}$ | $3.09 \times 10^{-3}$ | $-1.34 \times 10^{-3}$ | $3.36 \times 10^{-3} / 0.13 \times 10^{-3}$ | $2.01 \times 10^{-3}$ |
| $C_{17} + C_{19}$ | $2.98 \times 10^{-3}$ | $-1.42 \times 10^{-3}$ | $3.58 \times 10^{-3}/ 0.18 \times 10^{-3}$ | $1.34 \times 10^{-3}$ |
| $C_{14} + C_{16} + C_{18}$ | $2.29 \times 10^{-3}$ | $-1.05 \times 10^{-3}$ | $1.68 \times 10^{-3} / 0.13 \times 10^{-3}$ | $1.36 \times 10^{-3}$ |
| $C_{15} + C_{17} + C_{19}$ | $3.58 \times 10^{-3}$ | $-2.08 \times 10^{-3}$ | $3.12 \times 10^{-3} / 0.25 \times 10^{-3}$ | $1.92 \times 10^{-3}$ |
| $C_{15} + C_{19}$ | $3.76 \times 10^{-3}$ | $-2.32 \times 10^{-3}$ | $13.7 \times 10^{-3} / 1.12 \times 10^{-3}$ | $1.77 \times 10^{-3}$ |
| $C_{14}$-to-$C_{20}$ | $4.55 \times 10^{-3}$ | $-3.01 \times 10^{-3}$ | $5.37 \times 10^{-3} / 0.54 \times 10^{-3}$ | $2.89 \times 10^{-3}$ |

(1) The two values given for $\alpha_c$ are calculated using the two different slopes from Table 2 in the main text.

**Supporting Table S4. Experimentally measured unit cell parameters for rotator phases formed in the studied equimolar alkane mixtures.** The data is obtained in the experiments with emulsified samples at different temperatures.

| $T$, °C | Unit cell parameters | | | $T$, °C | Unit cell parameters | | |
|---|---|---|---|---|---|---|---|
| | $a$, Å | $b$, Å | $c/2$, Å | | $a$, Å | $b$, Å | $c/2$, Å |
| | $C_{16} + C_{17}$ | | | | $C_{14} + C_{16} + C_{18}$ | | |
| 18 | 7.781 | 4.995 | 23.076 | 18 | 7.815 | 4.964 | 23.851 |
| 16 | 7.747 | 5.005 | 23.052 | 16 | 7.779 | 4.984 | 23.824 |
| 12.5 | 7.685 | 5.024 | 23.057 | 12.5 | 7.715 | 5.013 | 23.810 |
| 10 | 7.642 | 5.035 | 23.045 | 10 | 7.652 | 5.027 | 23.801 |
| 8 | 7.612 | 5.042 | 23.036 | 8 | 7.633 | 5.031 | 23.797 |
| 4.5 | 7.564 | 5.051 | 23.034 | 4.5 | 7.569 | 5.046 | 23.790 |
| 2 | 7.537 | 5.056 | 23.032 | 2 | 7.548 | 5.050 | 23.787 |
| | $C_{15} + C_{17}$ | | | | $C_{15} + C_{17} + C_{19}$ | | |
| 14 | 7.862 | 4.949 | 22.769 | 18 | 8.027 | 4.859 | 24.692 |
| 13 | 7.846 | 4.952 | 22.661 | 16 | 8.013 | 4.881 | 24.662 |
| 12 | 7.825 | 4.964 | 22.633 | 12 | 7.940 | 4.901 | 24.110 |
| 10 | 7.790 | 4.980 | 22.559 | 10 | 7.817 | 4.950 | 24.051 |
| 7 | 7.715 | 5.000 | 22.513 | 7 | 7.732 | 4.981 | 24.034 |
| 4 | 7.652 | 5.022 | 22.509 | 4 | 7.657 | 5.008 | 24.037 |
| 2 | 7.628 | 5.033 | 22.508 | 2 | 7.632 | 5.017 | 24.015 |
| | $C_{16} + C_{18}$ | | | | $C_{15} + C_{19}$ (main mixed R) | | |
| 18 | 7.874 | 4.951 | 23.899 | 10 | 8.047 | 4.920 | 24.185 |
| 14 | 7.737 | 4.994 | 23.837 | 8 | 7.935 | 4.973 | 24.159 |
| 10 | 7.662 | 5.020 | 23.839 | 7 | 7.905 | 4.987 | 24.141 |
| 7 | 7.631 | 5.023 | 23.831 | 4 | 7.819 | 5.020 | 24.121 |
| 2 | 7.482 | 5.063 | 23.822 | 2 | 7.760 | 5.045 | 24.114 |
| | $C_{17} + C_{19}$ | | | | $C_{14}$-to-$C_{20}$ | | |
| 21 | 7.841 | 4.916 | 24.775 | 18 | 8.002 | 4.868 | 25.405 |
| 15 | 7.661 | 4.985 | 24.755 | 15 | 7.993 | 4.867 | 24.806 |
| 10 | 7.576 | 5.009 | 24.747 | 10 | 7.960 | 4.878 | 24.340 |
| 8 | 7.551 | 5.015 | 24.744 | 8 | 7.919 | 4.899 | 24.294 |
| 4 | 7.507 | 5.024 | 24.737 | 4 | 7.722 | 4.983 | 24.213 |
| 2 | 7.489 | 5.028 | 24.732 | 2 | 7.700 | 4.990 | 24.202 |

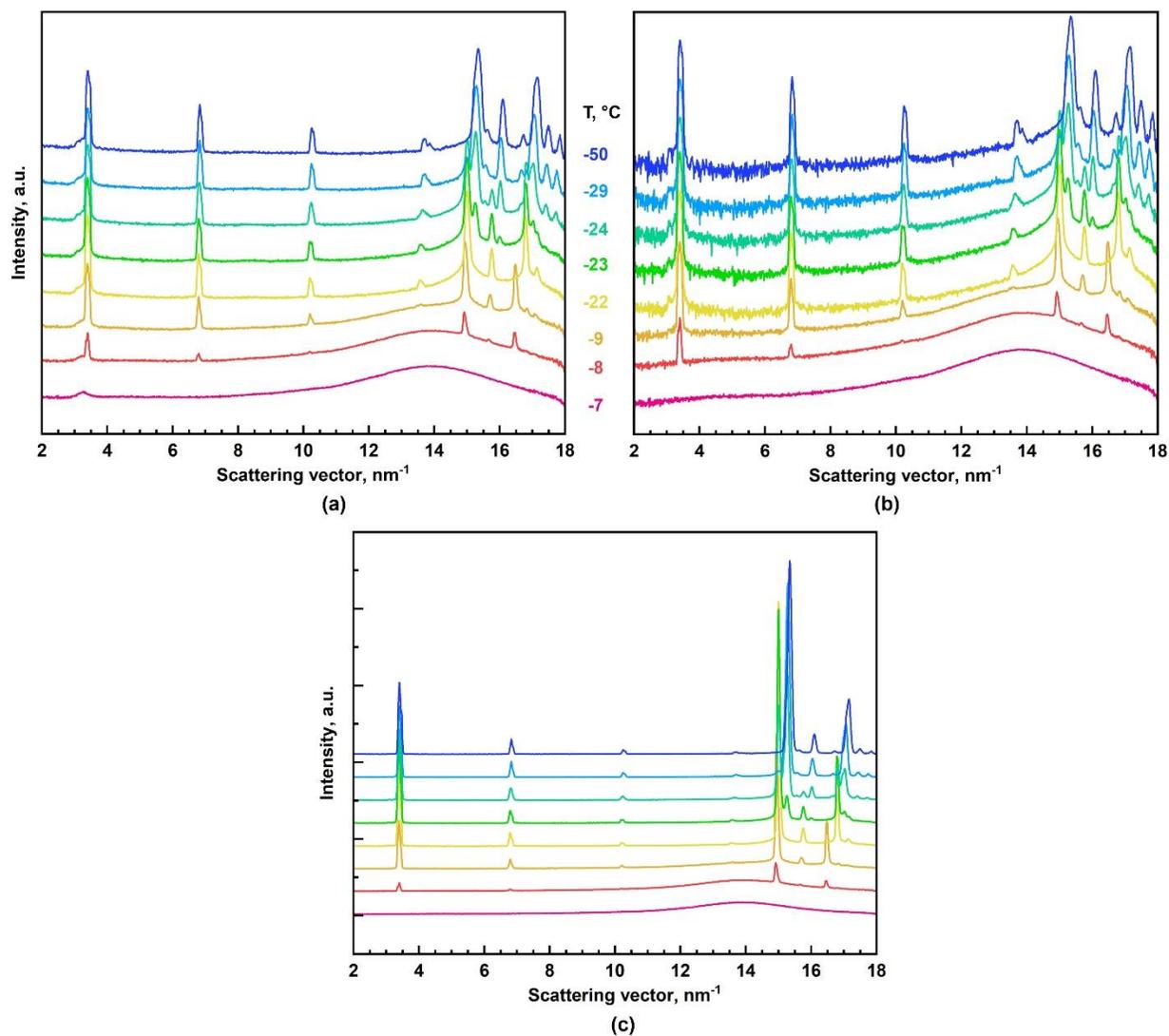

**Supporting Figure S1. SAXS/WAXS spectra of $C_{13}$ bulk alkane obtained upon cooling.** (a) Original data presented as *y*-axis in logarithmic scale. (b-c) The same signal as shown in (a) but with subtracted background. The background subtraction only increases the signal to noise ratio in the background without affecting the rest of the signal in any way. (b) shows the graph with *y*-axis in log scale and (c) in linear scale. The secondary and ternary reflections of the main (002) peak become much more pronounced when a logarithmic scale is employed.

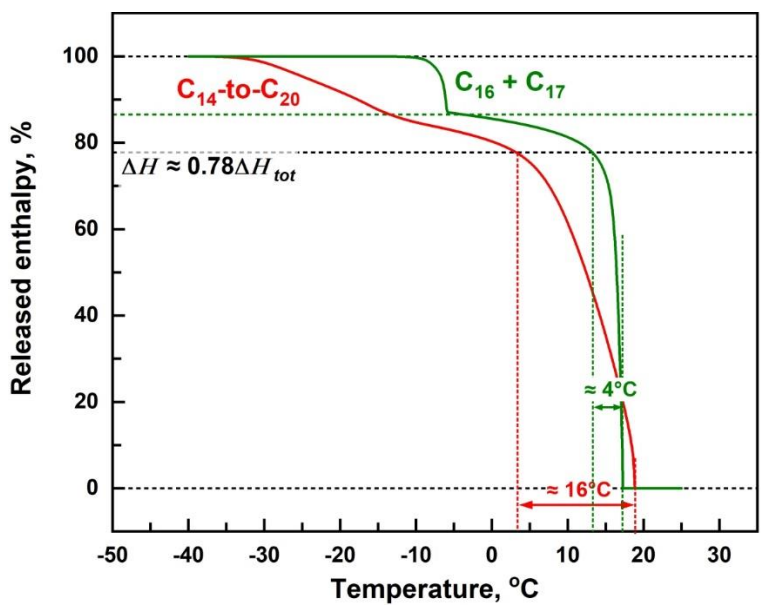

**Supporting Figure S2.** Comparison between the cumulative enthalpy released upon cooling of bulk $C_{14}$-to-$C_{20}$ mixture (red curve) and bulk $C_{16}$ + $C_{17}$ mixture (green curve). The temperature range in which about 78% of the total enthalpy is released upon cooling of the samples is ≈ 16°C for $C_{14}$-to-$C_{20}$ mixture, whereas it is just 4°C for $C_{16}$ + $C_{17}$ sample.

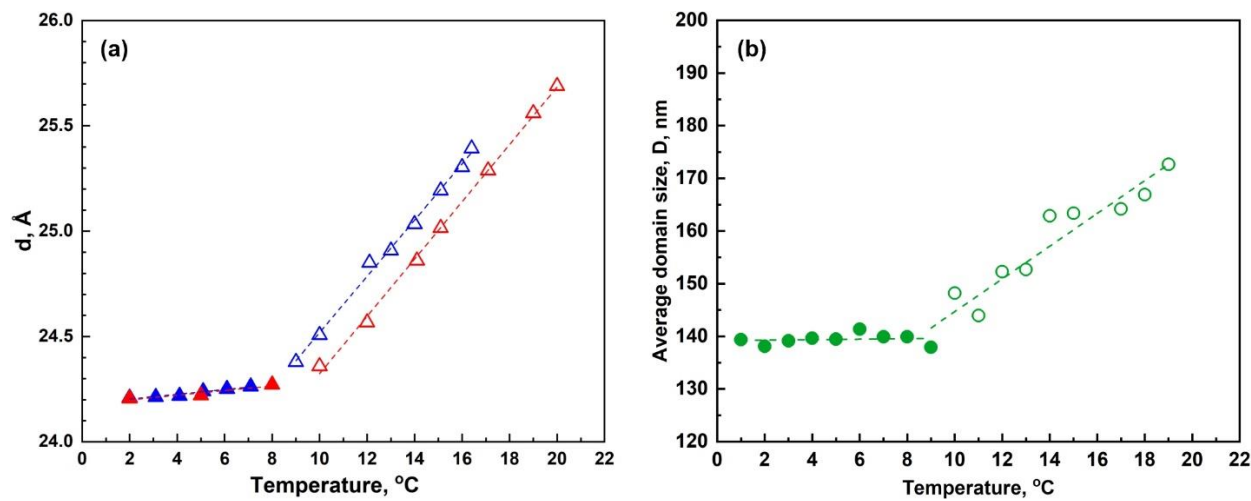

**Supporting Figure S3.** (a) Interlamellar spacing measured upon cooling (blue symbols) or heating (red symbols) for $C_{14}$-to-$C_{20}$ mixture dispersed as droplets in $C_{18}EO_{20}$ surfactant solution. The temperature dependence of the interlamellar spacing follows identical dependence upon both cooling and heating. The temperature derivatives are: $d(c/2)/dT \approx 0.011$ Å/°C for $T < 8$°C; $d(c/2)/dT \approx 0.133$ Å/°C upon cooling and $\approx 0.136$ Å/°C upon heating at $T > 8$°C. The difference in the interlamellar spacing measured at a given temperature $> 8$°C upon cooling and heating is attributed to the delay in rotator phase formation observed upon cooling. This delay arises due to the subcooling needed for rotator phase nucleus formation. (b) Average domain size calculated using Sherrer equation (constant K = 0.9, λ ≈ 1.54 Å) for the investigated $C_{14}$-to-$C_{20}$ emulsions.

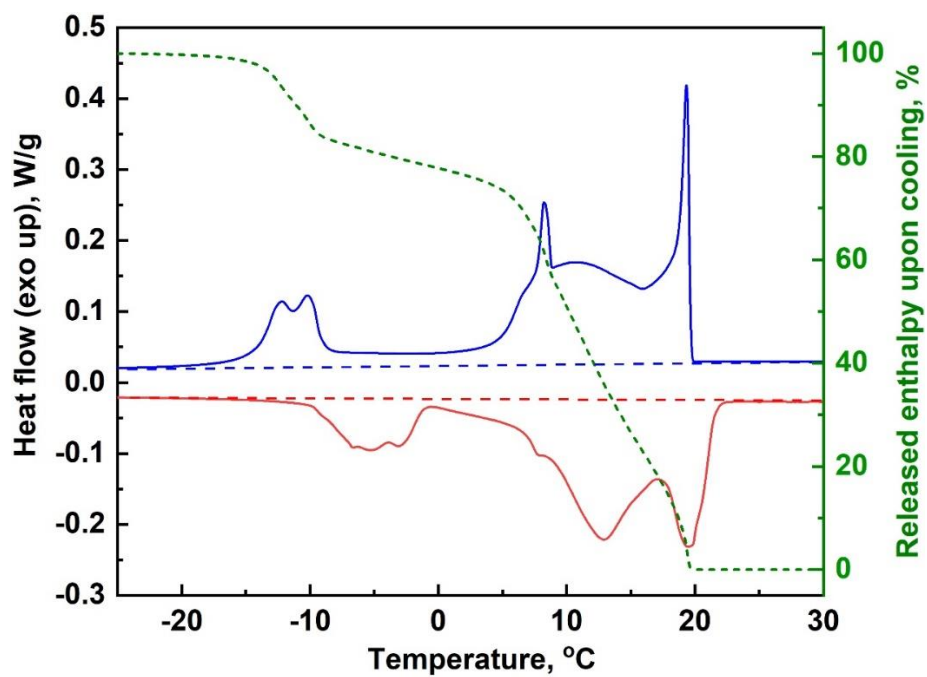

**Supporting Figure S4. DSC thermogram for $C_{15}$ + $C_{19}$ bulk alkane mixture.** The cooling signal is shown with the blue curve, while the heating signal is shown with red. The percentage area of the cooling peak as a function of temperature is shown with the dashed green line. Dashed blue and red lines show the baselines for the heat flow signals.

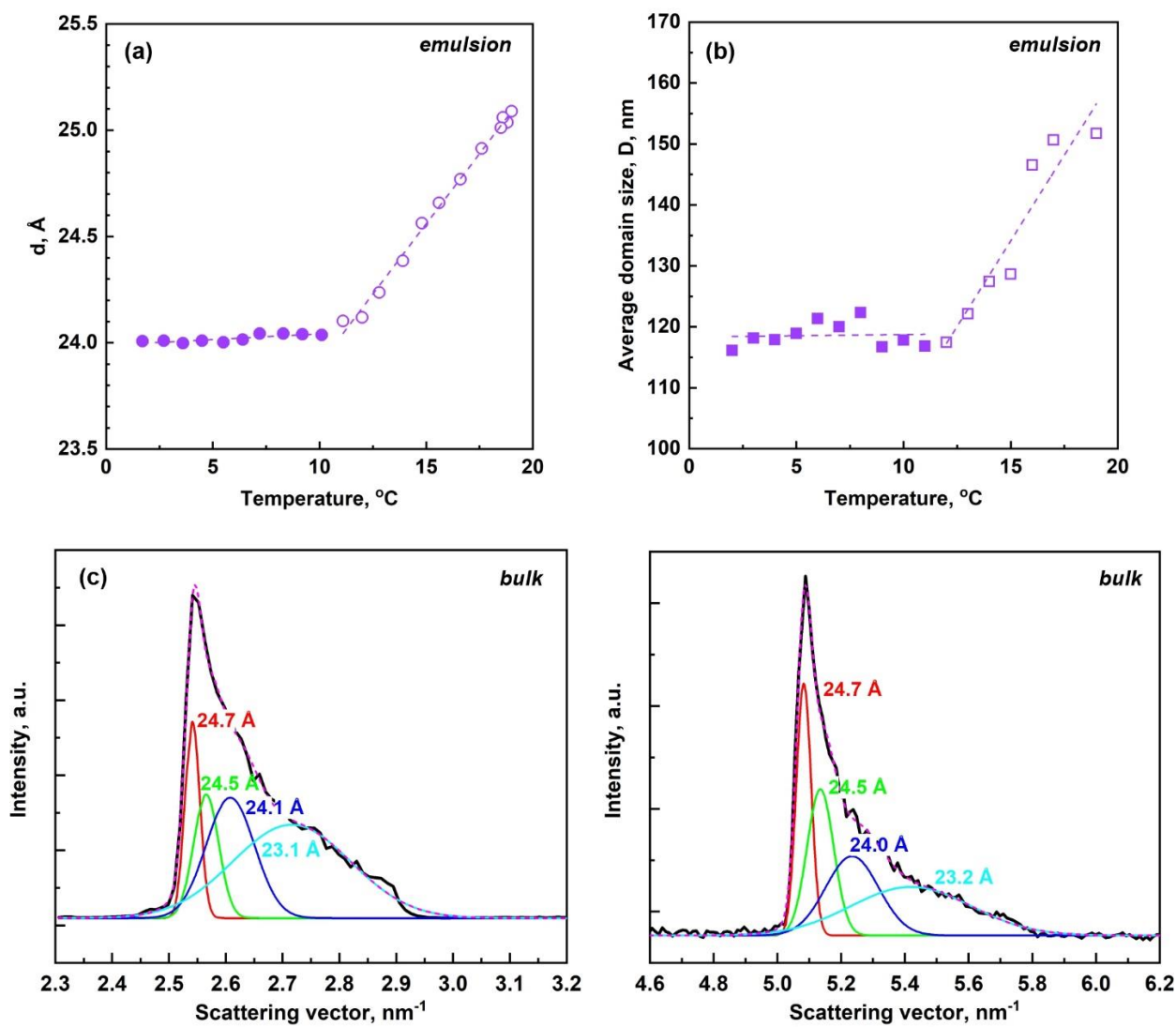

**Supporting Figure S5.** $C_{15}$ + $C_{17}$ + $C_{19}$ **mixture.** (a) Interlamellar spacing for ternary $C_{15}$ + $C_{17}$ + $C_{19}$ alkane mixture dispersed in $C_{16}EO_8$ surfactant. The data is from cooling experiment presented in Figure 7a in the main text. (b) Average crystallites domain size calculated using Sherrer equation (constant K = 0.9, λ ≈ 1.54 Å).(c) Peak deconvolution analysis for the SAXS spectra measured for bulk $C_{15}$ + $C_{17}$ + $C_{19}$ alkane mixture at $T$ = 2°C. Four different phases with similar *d*-values are needed to describe well the (002) and (004) peaks. The original signal is shown with black curve. The red, green, light blue and dark blue curves show the four separate Gaussian peaks used in the deconvolution analysis. The cumulative peak is shown with pink dashed line (which overlaps with the experimentally acquired signal shown in black).

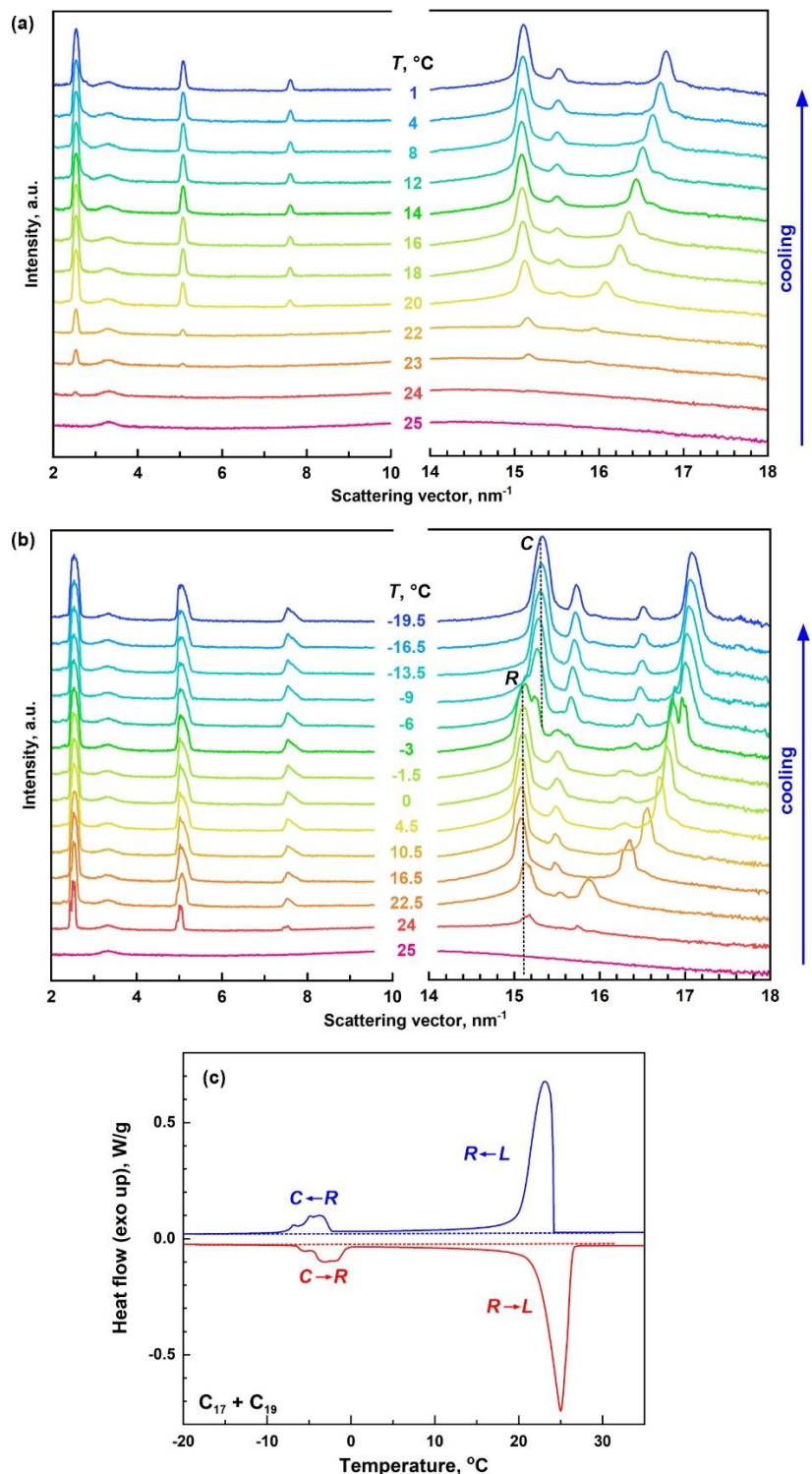

**Supporting Figure S6. Phase behavior of $C_{17}$ + $C_{19}$ binary mixture. (a,b)** SWAXS spectra obtained for: (a) emulsified in $C_{18}EO_{20}$ and (b) bulk sample. All molecules arrange in a single mixed rotator phase when emulsified, whereas several individual types of domains appear in the bulk mixture. **(c)** DSC thermogram.

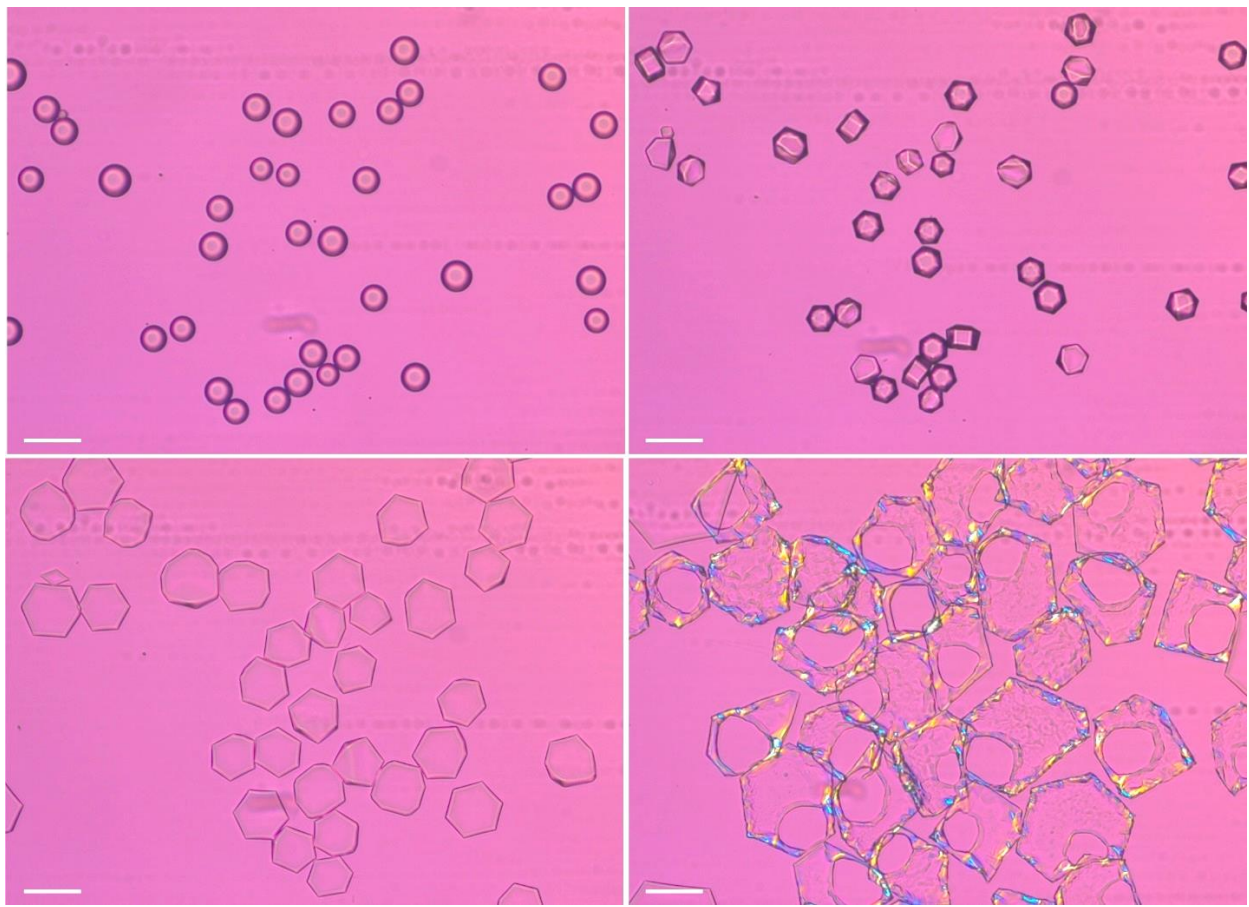

**Supporting Figure S7.** **Optical microscopy images obtained upon 0.75°C/min cooling for emulsion drops composed of $C_{15}$ + $C_{17}$ alkane mixture and dispersed in Tween 40 surfactant solution.** Spontaneous drop shape shifting is observed upon cooling. Details about the experimental set-up used to prepare these images are available in Refs [1,2]. Scale bars = 50 μm.

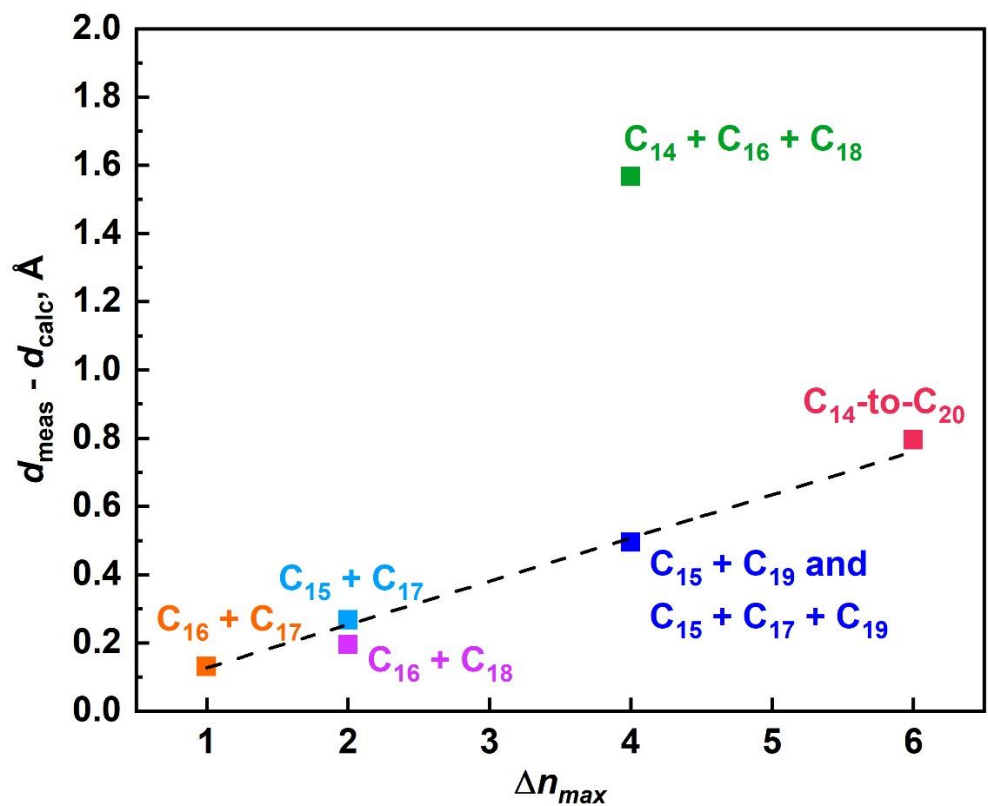

**Supporting Figure S8. Deviation of the rotator phase interlamellar spacing from its theoretical value.** Experimental data ($d_{meas}$) obtained at $T = 2°C$, when subtracted from the calculated interlamellar spacing using the equation $d_{calc} = 1.2724(\bar{n}-1) + 3.1476\,\text{Å}$, are presented relative to the maximal chain length difference in the studied alkane mixtures, $\Delta n_{max}$. A linear increase of this difference is observed with an increase in $\Delta n_{max}$ for all studied samples, except for $C_{14} + C_{16} + C_{18}$ mixture. In this specific case, the inclusion of the shorter $C_{14}$ alkane does not lead to decrease in the overall spacing in comparison to that measured for $C_{16} + C_{18}$ mixture. Instead, $\Delta n_{max}$ is increased, leading to this deviation. All remaining data are described by the linear relation: $d_{meas} - d_{calc} = 0.1268 \Delta n_{max}$.